
\input harvmac


\input epsf

\newcount\figno
\figno=0
\def\fig#1#2#3{
\par\begingroup\parindent=0pt\leftskip=1cm\rightskip=1cm\parindent=0pt
\baselineskip=11pt
\global\advance\figno by 1
\midinsert
\epsfxsize=#3
\centerline{\epsfbox{#2}}
\vskip 12pt
{\bf Fig. \the\figno: ~} #1\par
\endinsert\endgroup\par
}
\def\figlabel#1{\xdef#1{\the\figno}}
\def\frac#1#2{{#1 \over #2}}
\def\dg{{\sp\dagger}}
\def\Nfour{${\cal N} = 4$\ }
\def\de{\delta}
\def\De{\Delta}

\def\I{{\cal I}}

\def\O{{\cal O}}
\def\bO{{{\bar{\cal O}}}}
\def\Ob{{\cal O}^B}
\def\bOb{{\bar{\cal O}^B}}
\def\tO{{\tilde{\cal O}}}
\def\btO{{\bar{\tilde{\cal O}}}}

\def\PP{{\cal P}}
\def\K{{\cal K}}
\def\tpsi{{\tilde \psi}}
\def\tC{{\tilde C}}
\def\ta{{\tilde a}}

\def\CS{{\cal S}}
\def\x{x_1}
\def\xx{x_2}
\def\xxx{x_3}

\def\ep{\epsilon}

\def\tr{{\rm tr}}
\def\Tr{{\rm Tr}}
\def\hf{{1\over 2}}
\def\qu{{1\over 4}}
\def\ln{{\rm ln}}
\def\o{\over}
\def\til#1{\widetilde{#1}}

\def\b#1{\overline{#1}}

\def\lap{\Delta}
\def\cob{\delta}
\def\bra{\langle}
\def\ket{\rangle}
\def\lf{\left}
\def\ri{\right}
\def\riya{\rightarrow}

\def\la{\lambda}
\def\La{\Lambda}
\def\h#1{\widehat{#1}}

\def\bt{\beta}
\def\ga{\gamma}
\def\Ga{\Gamma}
\def\al{\alpha}
\def\bt{\beta}
\def\lam{\lambda}
\def\Lam{\Lambda}

\def\tens{\otimes}

\def\rt#1{\sqrt{#1}}

\def\sitarel#1#2{\mathrel{\mathop{\kern0pt #1}\limits_{#2}}}

\def\f{f^1_{23}}
\def\ff{f^2_{31}}
\def\fff{f^3_{12}}
\def\b{b_{12}}
\def\bb{b_{23}}
\def\bbb{b_{31}}

\lref\minzar{J.~A.~Minahan and K.~Zarembo, 
``The Bethe-ansatz for N = 4 super Yang-Mills,''
JHEP {\bf 0303}, 013 (2003)
[arXiv:hep-th/0212208].
}
\lref\ryzhov{
A.~V.~Ryzhov,
``Quarter BPS operators in N = 4 SYM,''
JHEP {\bf 0111}, 046 (2001)
[arXiv:hep-th/0109064].
}
\lref\dhoryz{
E.~D'Hoker and A.~V.~Ryzhov,
``Three-point functions of quarter BPS operators in N = 4 SYM,''
JHEP {\bf 0202}, 047 (2002)
[arXiv:hep-th/0109065].
}
\lref\dhofs{
E.~D'Hoker, D.~Z.~Freedman and W.~Skiba,
``Field theory tests for correlators in the AdS/CFT correspondence,''
Phys.\ Rev.\ D {\bf 59}, 045008 (1999)
[arXiv:hep-th/9807098].
}
\lref\dhofmmr{
E.~D'Hoker, D.~Z.~Freedman, S.~D.~Mathur, A.~Matusis and L.~Rastelli,
``Extremal correlators in the AdS/CFT correspondence,''
[arXiv:hep-th/9908160].
}
\lref\biakrs{
M.~Bianchi, S.~Kovacs, G.~Rossi and Y.~S.~Stanev,
``Properties of the Konishi multiplet in N = 4 SYM theory,''
JHEP {\bf 0105}, 042 (2001)
[arXiv:hep-th/0104016].
}
\lref\chukt{
C.~S.~Chu, V.~V.~Khoze and G.~Travaglini,
``Three-point functions in N = 4 Yang-Mills theory and pp-waves,''
JHEP {\bf 0206}, 011 (2002)
[arXiv:hep-th/0206005].
}
\lref\confhmmps{
N.~R.~Constable, D.~Z.~Freedman, M.~Headrick, S.~Minwalla, L.~Motl, A.~Postnikov and W.~Skiba,
``PP-wave string interactions from perturbative Yang-Mills theory,''
JHEP {\bf 0207}, 017 (2002)
[arXiv:hep-th/0205089].
}
\lref\GrossMH{
D.~J.~Gross, A.~Mikhailov and R.~Roiban,
``A calculation of the plane wave string Hamiltonian from N = 4 super-Yang-Mills theory,''
JHEP {\bf 0305}, 025 (2003)
[arXiv:hep-th/0208231].
}

\lref\ConstableVQ{
N.~R.~Constable, D.~Z.~Freedman, M.~Headrick and S.~Minwalla,
``Operator mixing and the BMN correspondence,''
JHEP {\bf 0210}, 068 (2002)
[arXiv:hep-th/0209002].
}

\lref\aruppss{
G.~Arutyunov, S.~Penati, A.~C.~Petkou, A.~Santambrogio and E.~Sokatchev,
``Non-protected operators in N = 4 SYM and multiparticle states of AdS(5)
SUGRA,''
Nucl.\ Phys.\ B {\bf 643}, 49 (2002)
[arXiv:hep-th/0206020].
}
\lref\beikpss{
N.~Beisert, C.~Kristjansen, J.~Plefka, G.~W.~Semenoff and M.~Staudacher,
``BMN correlators and operator mixing in N = 4 super Yang-Mills theory,''
Nucl.\ Phys.\ B {\bf 650}, 125 (2003)
[arXiv:hep-th/0208178].
}
\lref\beiks{
N.~Beisert, C.~Kristjansen and M.~Staudacher,
``The dilatation operator of N = 4 super Yang-Mills theory,''
Nucl.\ Phys.\ B {\bf 664}, 131 (2003)
[arXiv:hep-th/0303060].
}
\lref\bei{
N.~Beisert,
``The complete one-loop dilatation operator of N = 4 super Yang-Mills
theory,''
Nucl.\ Phys.\ B {\bf 676}, 3 (2004)
[arXiv:hep-th/0307015].
}
\lref\beis{
N.~Beisert and M.~Staudacher,
``The N = 4 SYM integrable super spin chain,''
Nucl.\ Phys.\ B {\bf 670}, 439 (2003)
[arXiv:hep-th/0307042].
}
\lref\tse{
A.~A.~Tseytlin,
``Spinning strings and AdS/CFT duality,''
[arXiv:hep-th/0311139].
}
\lref\frejmv{
D.~Z.~Freedman, K.~Johnson, R.~Munoz-Tapia and X.~Vilasis-Cardona,
``A cutoff procedure and counterterms for differential renormalization,''
Nucl.\ Phys.\ B {\bf 395}, 454 (1993)
[arXiv:hep-th/9206028];~
D.~Z.~Freedman, K.~Johnson and J.~I.~Latorre,
``Differential regularization and renormalization: a new method of calculation
in quantum field theory,''
Nucl.\ Phys.\ B {\bf 371}, 353 (1992).
}

\lref\gomrs{
C.~Gomez, M.~Ruiz-Altaba, and G.~Sierra, {\it Quantum Groups in Two
Dimensional Physics}, Cambridge University Press, 1996.}

\lref\benrp{
I.~Bena, J.~Polchinski and R.~Roiban,
``Hidden symmetries of the AdS(5) x S**5 superstring,''
Phys.\ Rev.\ D {\bf 69}, 046002 (2004)
[arXiv:hep-th/0305116].
}
\lref\vall{
B.~C.~Vallilo,
``Flat currents in the classical AdS(5) x S**5 pure spinor superstring,''
[arXiv:hep-th/0307018].
}
\lref\cheww{
B.~Chen, X.~J.~Wang and Y.~S.~Wu,
``Integrable open spin chain in super Yang-Mills and the plane-wave / SYM
duality,''
JHEP {\bf 0402}, 029 (2004)
[arXiv:hep-th/0401016];
B.~Chen, X.~J.~Wang and Y.~S.~Wu,
``Open spin chain and open spinning string,''
[arXiv:hep-th/0403004].
}
\lref\dewman{
O.~DeWolfe and N.~Mann,
``Integrable open spin chains in defect conformal field theory,''
[arXiv:hep-th/0401041].
}
\lref\sklyan{
E.~K.~Sklyanin,
``Boundary conditions for integrable quantum systems,''
J.\ Phys.\ A {\bf 21}, 2375 (1988).
}
\lref\stefan{
B.~Stefanski~jr.,
``Open spinning strings,''
JHEP {\bf 0403}, 057 (2004)
[arXiv:hep-th/0312091].
}
\lref\BMN{
D.~Berenstein, J.~M.~Maldacena and H.~Nastase,
``Strings in flat space and pp waves from N = 4 super Yang Mills,''
JHEP {\bf 0204}, 013 (2002)
[arXiv:hep-th/0202021].
}
\lref\SpradlinAR{
M.~Spradlin and A.~Volovich,
``Superstring interactions in a pp-wave background,''
Phys.\ Rev.\ D {\bf 66}, 086004 (2002)
[arXiv:hep-th/0204146].
}
\lref\Kugo{
T.~Kugo, H.~Kunitomo and K.~Suehiro,
``Nonpolynomial closed string field theory,''
Phys.\ Lett.\ B {\bf 226}, 48 (1989).
}
\lref\Saadi{
M.~Saadi and B.~Zwiebach,
``Closed string field theory from polyhedra,''
Annals Phys.\  {\bf 192}, 213 (1989).
}
\lref\BeisertTN{
N.~Beisert,
``BMN operators and superconformal symmetry,''
Nucl.\ Phys.\ B {\bf 659}, 79 (2003)
[arXiv:hep-th/0211032].
}
\lref\BeisertTQ{
N.~Beisert, C.~Kristjansen and M.~Staudacher,
``The dilatation operator of N = 4 super Yang-Mills theory,''
Nucl.\ Phys.\ B {\bf 664}, 131 (2003)
[arXiv:hep-th/0303060].
}
\lref\WittenSFT{
E.~Witten,
``Noncommutative geometry and string field theory,''
Nucl.\ Phys.\ B {\bf 268}, 253 (1986).
}
\lref\GiddingsBP{
S.~B.~Giddings and E.~J.~Martinec,
``Conformal geometry and string field theory,''
Nucl.\ Phys.\ B {\bf 278}, 91 (1986).
}

\lref\BeisertYS{
N.~Beisert,
``The su(2$|$3) dynamic spin chain,''
arXiv:hep-th/0310252.
}
\lref\BianchiEG{
M.~Bianchi, G.~Rossi and Y.~S.~Stanev,
``Surprises from the resolution of operator mixing in N = 4 SYM,''
Nucl.\ Phys.\ B {\bf 685}, 65 (2004)
[arXiv:hep-th/0312228].
}
\lref\leemrs{
S.~M.~Lee, S.~Minwalla, M.~Rangamani and N.~Seiberg,
``Three-point functions of chiral operators in D = 4, N = 4 SYM at  large N,''
Adv.\ Theor.\ Math.\ Phys.\  {\bf 2}, 697 (1998)
[arXiv:hep-th/9806074].
}
\lref\dolnw{
L.~Dolan, C.~R.~Nappi and E.~Witten,
``A relation between approaches to integrability in superconformal Yang-Mills
theory,''
JHEP {\bf 0310}, 017 (2003)
[arXiv:hep-th/0308089].
}
\lref\alday{
L.~F.~Alday,
``Nonlocal charges on AdS(5) x S**5 and pp-waves,''
JHEP {\bf 0312}, 033 (2003)
[arXiv:hep-th/0310146].
}


\Title{\vbox{\baselineskip12pt
\hbox{hep-th/0404190}
\hbox{EFI-04-14}
\vskip-.5in}}
{\vbox{\centerline{Three-Point Functions in \Nfour SYM Theory at One-Loop}}}
\centerline{Kazumi Okuyama$^1$
and$\,~$Li-Sheng Tseng$^{2,3}$
}
\bigskip\medskip
\centerline{${}^1$\it Enrico Fermi Institute, University of Chicago} 
\centerline{\it  5640 S. Ellis Ave., Chicago, IL 60637, USA}
\centerline{\tt  kazumi@theory.uchicago.edu}
\medskip
\centerline{${}^2$\it Enrico Fermi Institute and Department of Physics}
\centerline{\it  University of Chicago, Chicago, IL 60637, USA}
\medskip
\centerline{${}^3$\it Department of Physics, University of Utah}
\centerline{\it Salt Lake City, UT 84112, USA}
\centerline{\tt  tseng@physics.utah.edu}
\baselineskip18pt
\medskip\bigskip\medskip\bigskip\medskip
\baselineskip16pt

\noindent
We analyze the one-loop correction to the three-point function
coefficient of scalar primary operators in \Nfour SYM theory.
By applying constraints from the superconformal symmetry, we demonstrate
that the type of Feynman diagrams that contribute depends on the 
choice of renormalization scheme.   In the planar limit, explicit
expressions for the correction are interpreted in terms of the
hamiltonians of the associated integrable closed and open spin chains.
This suggests that at least at one-loop, the planar conformal field
theory is integrable with the anomalous dimensions and OPE
coefficients both obtainable from integrable spin chain calculations.
We also connect the planar results with similar structures found in
closed string field theory.

\Date{April, 2004}


\newsec{Introduction}

Integrability in quantum field theory has for the most part been
relegated to the realms of two dimensional theories.  (For an
overview, see for example, \gomrs).  Recently, there has been much
excitement over the prospects of integrability in four dimensional
gauge theory.  Much of the recent work has focused on \Nfour super Yang-Mills
(SYM) theory.  This theory is special in that it is superconformal and is
widely believed to be dual to type IIB string theory on $AdS_5\times S^5$ by
the AdS/CFT correspondence.

In the large $N$ planar limit of \Nfour SYM, an important
observation was put forth by Minahan and Zarembo \minzar\ concerning
the first order $\lam=g^2_{\rm YM} N$ correction
to the scaling dimension of composite single trace operators
consisting of derivative-free scalar fields.  They pointed out that
these operators can be naturally mapped into states of an integrable
$SO(6)$ spin chain.  And amazingly, the hamiltonian of the spin chain
is proportional to the $O(\lam)$ correction to the scaling dimension of the
operators.  In follow up works, Beisert, {\it et al.} \refs{\beiks,\bei,\beis}
extended this relation to arbitrary single trace composite operators
and showed that the first order correction to the dilation operator is
in fact given by the hamiltonian of the integrable $PSU(2,2|4)$ spin chain.

Further hints of \Nfour SYM planar integrability has also emerged
via the AdS/CFT correspondence.  Type IIB theory on $AdS_5\times S^5$
has been found to contain an infinite number of nonlocal charges
\refs{\benrp,\vall,\alday}.\foot{A discussion of nonlocal charges in
free \Nfour SYM theory can be found in \dolnw.}  And in the extensive
analysis of spinning strings solutions, integrable structures have played a
central role (see \tse\ for a review).  Yet, with all these
suggestions, the origin of this apparent planar integrability in \Nfour
SYM theory is still not clear. 

So far, evidences of planar integrability have all been gathered from analyses
of the scaling dimension spectrum of operators in the theory.  
But if the planar \Nfour SYM theory is indeed integrable, then the
dynamical aspects of the theory must also be describable by
integrable structures.  For a conformal field theory, the dynamical
data are encoded in the coefficients of the operator product
expansions (OPE), i.e. the structure constants.  These structure
constants, directly related to the coefficient of the three-point
functions, are important for solving the correlation functions of the
theory.

In this paper, we analyze the one-loop correction to
the three-point function coefficients of conformal primary
operators.  We will neglect possible additional $O(\lam)$ correction
from $\lam$ dependent operator mixing that can only be seen at the
two-loop level \aruppss.\foot{A discussion of these additional
contributions is given in section 7.}   As with Minahan and
Zarembo, we will consider only
scalar conformal primaries in the $SO(6)$ subsector of the theory and
emphasize the planar limit.\foot{Some previous calculations of
non-protected \Nfour SYM scalar three-point functions can be found in
\refs{\beikpss,\confhmmps,\chukt}.}  Our calculations are much
simplified by exploiting 
constraints from the superconformal symmetry of the theory.  As is
known from the study of non-renormalization of two- and three-point
functions of BPS operators, \Nfour
supersymmetry relates functional dependence of various $O(\lam)$ Feynman
diagrams \refs{\dhofs}.  Imposing further the constraints from
conformal invariance, the first order correction to the three-point
coefficients can be obtained, surprisingly, without summing all
possible Feynman diagrams.  As we will show, the first order
correction may be obtained just by summing up only two-point Feynman
diagram interactions (i.e. diagrams that connect two
operators).\foot{This has been utilized in calculations of three-point
functions of BPS operators in \dhoryz.}  Alternatively, one can choose
to sum up mainly three-point Feynman diagram interactions.  These two
prescriptions are just the result of working in two special
renormalization schemes.  Since the physical correction must be scheme
independent, the two prescriptions  provide us with two different, but
complementary, pictures of the three-point function interaction.

In the planar limit, the two-point Feynman diagram description leads
naturally to a spin chain interpretation.  For the generic non-extremal
three-point function, the one-loop correction to the three-point
coefficient turns out to involve the hamiltonian of both
closed and open $SO(6)$ integrable spin chains.  Open spin chains can
arise by splitting a periodic closed spin chain into two
``correlated'' open spin chains.  As we will show, the three primary
operators can be combined to give
three open spin chain density matrices.  Indeed, the one-loop correction
contains a contribution from the ensemble average of the open spin chain
hamiltonian with respect to the three density matrices.  As for the
three-point Feynman diagram description, it does not seem to have 
a direct interpretation using conventional spin chain language.  Its natural
setting seems to be that of string field theory.  We shall identify
the planar three-point Feynman diagram prescription with an analogous
structure, the Witten type three-string vertex, in covariant 
closed string field theory.

In section 2, we provide the general form of two- and
three-point functions of conformal primary operators at one-loop.  In
section 3, we first explain how supersymmetry relates various
Feynman diagrams.  Although our presentation will mainly focus on the
planar limit, we will utilize and point out results that are valid at
finite $N$.  We proceed to apply the conformal symmetry constraints to
calculate the three-point function coefficient.  Of interest is that
the one-loop correction of the extremal three-point function
coefficient depends only on the anomalous dimensions.  In section 4,
we provide explicit results in the planar limit.  The planar results
are interpreted in the context of integrable spin
chains in section 5.  In section 6, from our SYM results, we infer
some properties of closed string field theory in $AdS_5\times S^5$.  We
close with a discussion in section 7.  Appendix A contains the
computation of integrals that appear in the contributing Feynman
diagrams.  And in appendix B, we give a matrix integral representation
of $SO(6)$ index structure of three-point functions.

\newsec{General Form of Two-Point and Three-Point Function at One-Loop}

Let $\Ob$ denote bare scalar conformal primary operators.  For
bare operators with identical free scaling dimension $\De_0$, the
two-point function to first order in $\lam=g^2_{\rm YM} N$ takes the form, 
\eqn\twobare{\eqalign{\bra\bOb_\al(x_1)\Ob_\bt(x_2)\ket
&=\frac{1}{|x_{12}|^{2\De_0}} \left [ g_{\al\bt}- h_{\al\bt}\,
\ln|x_{12}\Lam|^2 \right ] \cr &=\frac{1}{|x_{12}|^{2\De_0}} g_{\al\rho}
\left [ \de^\rho_\bt - g^{\rho\sigma} h_{\sigma\bt}
\,\ln|x_{12}\Lam|^2 \right ]} }
where $x_{12}^\mu = x_1^\mu - x_2^\mu$, $\;(g^{-1}h)^\al_{~\bt}$ is the
anomalous dimension matrix, and $\De_{0\al}=\De_{0\bt}=\De_0$.  Note
that both $g_{\al\bt}$ and $h_{\al\bt}$ may
contain terms of $O(\lam)$.  By a linear
transformation, $\Ob_\al \to \O_\al=\Ob_\rho\,U^{\rho}_{~\al}$, the
matrices can be diagonalized as follows \refs{\GrossMH,\ConstableVQ},
\eqn\diagmat{\eqalign{(U^\dg)_\al^{~\rho}\, g_{\rho\sigma}\,
U^\sigma_{~\bt} &=\; \delta_{\al\bt}\,A_\al \cr
(U^{-1})\,\!^\al_{~\rho}\, (g^{-1}h)^\rho_{~\sigma}\, 
U^\sigma_{~\bt} &=\; \delta^\al_{~\bt}\,\ga_\al~.} }
Here, $A_\al$ and
$\ga_\al$ are the normalizations and anomalous dimensions,
respectively.  Notice that at the one-loop level, $U^{\rho}_{~\al}$ is
independent of $\lam$; therefore, $\lam$ dependent operator mixing
does not appear at one-loop.  We will decompose the normalization as
$A_\al = N^2_\al\,[1+ 2 a_\al\,\lam +O(\lam^2) ]$, where $a_\al$ is a
scheme dependent constant.  The two-point function for an
orthogonalized operator (or eigen-operator) $\O_\al$ then becomes
\eqn\twobare{\eqalign{\bra\bO_\al(x_1)\O_\al(x_2)\ket
&=\frac{N_\al^2 [1+ 2 a_\al\,\lam]}{|x_{12}|^{2\De_0}}\left [ 1- \ga_\al\,
\ln|x_{12}\Lam|^2 \right ] \cr &= \frac{N_\al^2}{|x_{12}|^{2\De_0}}
\left [1+ 2 a_\al\,\lam - \ga_\al\, \ln|x_{12}\Lam|^2 \right ] } }

From the bare eigen-operator, the renormalized operator is
defined to be
\eqn\renormdef{
\tO_\al=\O_\al\; [1- a_\al\,\lam + \ga_\al\;\ln|\Lam/\mu| + O(\lam^2)]}
where $\mu$ is the renormalization scale.  As required by conformal
invariance, the two-point function of renormalized primary operators
takes the form,
\eqn\tworen{\bra\btO_\al(x_1)\tO_\al(x_2)\ket =
\frac{N^2_\al}{|x_{12}|^{2\De_{0\al}} |x_{12}\,\mu|^{2\ga_\al}}}
with scaling dimension $\De_\al=\De_{0\al}+\,\ga_\al$.  Note that $\tO_\al$ is
conventionally defined such that $N_\al=1$.  For convenience, we will
only require that the renormalized operator be orthogonalized and not
orthonormalized.  

Conformal invariance constrains the three-point function for
renormalized primary operators to be
\eqn\threeren{\bra\btO_\al(x_1)\tO_\bt(x_2)\tO_\rho(x_3)\ket =
\frac{N_\al N_\bt N_\rho\;
c_{\al\bt\rho}}{|x_{12}|^{\De_\al+\De_\bt-\De_\rho}|x_{13}|^{\De_\al+\De_\rho-\De_\bt}|x_{23}|^{\De_\bt+\De_\rho-\De_\al}\,|\mu|^{\ga_\al+\ga_\bt+\ga_\rho}}}
where $c_{\al\bt\rho}$ is the three-point function coefficient.  We are
interested in finding the one-loop correction to the structure
constant.  Therefore, we decompose 
\eqn\conedef{c_{\al\bt\rho}=c^0_{\al\bt\rho}\Big(1
+ \lam\; c^1_{\al\bt\rho} + O(\lam^2)\Big)
}
assuming $c^0_{\al\bt\rho}\neq 0$.
Substituting \renormdef\ into \threeren , the three-point function for
bare eigen-operators is given at one-loop by 
\eqn\threebare{\eqalign{\bra\bO_\al(x_1)\O_\bt(x_2)\O_\rho(x_3)\ket
& = \frac{N_\al N_\bt N_\rho\;c^0_{\al\bt\rho}(1
+ \lam\;
c^1_{\al\bt\rho})\;\left[ 1+\lam (a_\al+a_\bt+a_\rho)\right]}
{|x_{12}|^{\De_\al+\De_\bt-\De_\rho}|x_{13}|^{\De_\al+\De_\rho-\De_\bt}|x_{23}|^{\De_\bt+\De_\rho-\De_\al}\,|\Lam|^{\ga_\al+\ga_\bt+\ga_\rho}}\cr   
&=\frac{C^0}{|x_{12}|^{\De_{0\al}+\De_{0\bt}-\De_{0\rho}}|x_{13}|^{\De_{0\al}+\De_{0\rho}-\De_{0\bt}}|x_{23}|^{\De_{0\bt}+\De_{0\rho}-\De_{0\al}}}
\times \cr  & \qquad \times \Big( 1 + \lam \;C^1\cr &\qquad\qquad -\ga_\al\,{\ln|\frac{x_{12}x_{13}\Lam}{x_{23}}|}
-\ga_\bt\,{\ln|\frac{x_{12}x_{23}\Lam}{x_{13}}|}-\ga_\rho\,{\ln|\frac{x_{23}x_{13}\Lam}{x_{12}}|}\;\Big) }}
where we have defined $C^0=N_\al N_\bt
N_\rho c^0_{\al\bt\rho}$ as the overall factor and $C^1=c^1_{\al\bt\rho} +
a_\al+a_\bt+a_\rho$ as the constant one-loop correction.  Notice that
$C^1$ is dependent on the renormalization scheme.  However, we are
interested in calculating $c^1_{\al\bt\rho}=C^1- a_\al-a_\bt-a_\rho$
which is scheme independent.  

If desired, we can transform the three-point expression in \threebare\
into that for bare operators, $\Ob$, by applying the linear
transformation, $\O_\al=\Ob_\rho\,U^{\rho}_{~\al}$.

\newsec{Perturbative Calculations of Three-Point Functions}

We write the Euclidean \Nfour action with $SU(N)$ gauge symmetry as 
\eqn\action{\CS=\frac{N}{\lam}\int d^4x\; {\rm tr}\left\{\frac{1}{2}
\,F_{\mu\nu}F^{\mu\nu}+ D^{\mu}\phi^i D_{\mu}\phi^i -
\frac{1}{2}\,[\phi^i,\phi^j]^2 + {\rm fermions~terms} \right\}~}
with the scalar fields $\phi^i$, $i=1,\ldots,6$, in the {\bf 6} of the
$SO(6)$ ($R$ symmetry group).  We will only consider conformal primary
operators composed of derivative-free scalar fields.  At
finite $N$, the operators of interest must belong to the closed $SU(2)$
subsector of the theory \refs{\beiks,\bei}.  These are operators that
can be expressed in terms of only two complex scalar fields, for example,
$Z=\phi^1+i\phi^2$ and $W=\phi^3+i\phi^4$.  By $SO(6)$ representation
theory arguments, such operators do not mix with
scalar combinations of fermion bilinears, field strengths
$F_{\mu\nu}$, or covariant derivatives $D_\mu$ \beiks.  On the other
hand, when working in the planar limit, we will choose to focus on the single
trace operators in the $SO(6)$ subsector.  In the planar limit, single trace
operators do not mix with multi-trace operators.  This allows us to
neglect subtleties that may arise due to the regularization of
multi-trace operators \BianchiEG\ in the $SO(6)$ sector.  We will
write $SO(6)$ single trace operators as
\eqn\opdef{\O[\psi_I]=\frac{1}{\lam^{L/2}}\,\psi_{i_1\ldots i_L}
\tr\,\phi^{i_1}\cdots\phi^{i_L}} 
where $I=\{i_1,\ldots,i_L\}$ and the constant coefficients
$\psi_{i_1\ldots i_L}$ are independent of $N$ and $\lam$.   The
additional $\lam^{-L/2}$ factor in \opdef\ is inserted so that
the planar two-point functions $\bra\bO[\psi_I]\O[\psi_J]\ket\sim
O(N^0)$ and the planar three-point functions
$\bra\bO[\psi_I]\O[\psi_J]\O[\psi_K]\ket \sim O(N^{-1})$.

Below, we will first review some important characteristics of the
functional forms of Feynman diagrams at one-loop found in the studies
of two- and three-point functions of BPS operators
\refs{\dhofs,\ryzhov,\dhoryz}.  Since these relations are independent
of $N$, we will discuss the Feynman diagrams below in the planar
limit.  The planar Feynman diagrams will also play an important role
in section 4.  We will then utilize conformal symmetry to obtain
general formulas at finite $N$ for $c^1_{\al\bt\rho}$, the one-loop
correction to the three-point function coefficient defined in \conedef.

\subsec{Feynman Diagrams in the Planar Limit}

For three-point functions, two classes of Feynman diagrams contribute.
Contributions that arise already in the two-point function are shown
in Fig.\ 1.  In the planar limit, these interactions can at most be
nearest-neighbor.  Acting on two operators
$\O[\psi_I](\x)\O[\psi_J](\xx)$, the diagrams are expressed with
emphasis on the $SO(6)$ indices as follows. 
\fig{Two-point Feynman diagrams.}{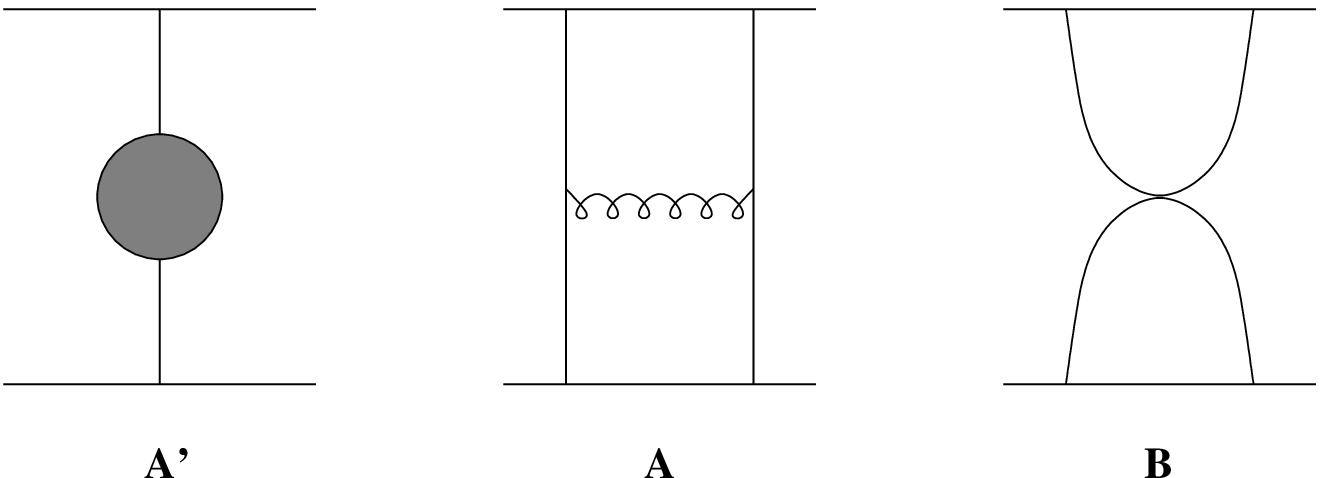}{3.3in}
\eqn\twofeyn{\eqalign{{\rm Diagram\
A'}&=\;\de^{j_l}_{i_l}\,A'(\x,\xx)\,N^{-1}\, G(\x,\xx) \cr
{\rm Diagram\ A}~&=\;\de^{j_l}_{i_l}\de^{j_{l+1}}_{i_{l+1}}\,A(\x,\xx)\, N^{-1}\,[G(\x,\xx)]^2 \cr
{\rm Diagram\ B}~&=\left (2\de^{j_{l+1}}_{i_l}\de^{j_{l}}_
{i_{l+1}} - \de^{j_l}_{i_l}\de^{j_{l+1}}_{i_{l+1}} - \de_{i_l
i_{l+1}}\de^{j_{l}j_{l+1}} \right )\, B(\x,\xx)\, N^{-1}\,[G(\x,\xx)]^2} }
where we have extracted factors of $1/N$ and the free scalar propagator
$G(\x,\xx)=\frac{\lam}{N}\,\frac{1}{8\pi^2|x_{12}|^2}$ in
defining the functions, $A'(\x,\xx)$, $A(\x,\xx)$, and
$B(\x,\xx)$.\foot{The two-point Feynman diagram arising from the
potential
$\tr\,[\phi^i,\phi^j]^2$ and having three contractions with
$\O[\psi_I](\x)$ and one contraction with $\O[\psi_J](\xx)$ must have
a zero net contribution.  Besides being quadratically divergent, a
non-zero contribution for this diagram would imply
perturbative mixing between primary operators of different free
scaling dimensions.  But such mixing is prohibited by conformal symmetry.}
Exact forms of these functions can be calculated directly.  However, a
relation between these functions is provided by the
non-renormalization theorem of correlators of BPS
operators \dhofs.  Consider the $\frac{1}{2}$ BPS 
operator $\tr\,\phi^1\phi^2(x)$.  Non-renormalization of $\bra
\tr\,\phi^1\phi^2(\x) \tr\,\phi^1\phi^2(\xx) \ket$ implies
\eqn\twononren{A'(\x,\xx)+A(\x,\xx)+B(\x,\xx)=0~.}
\noindent 

\fig{Three-point Feynman diagrams.}{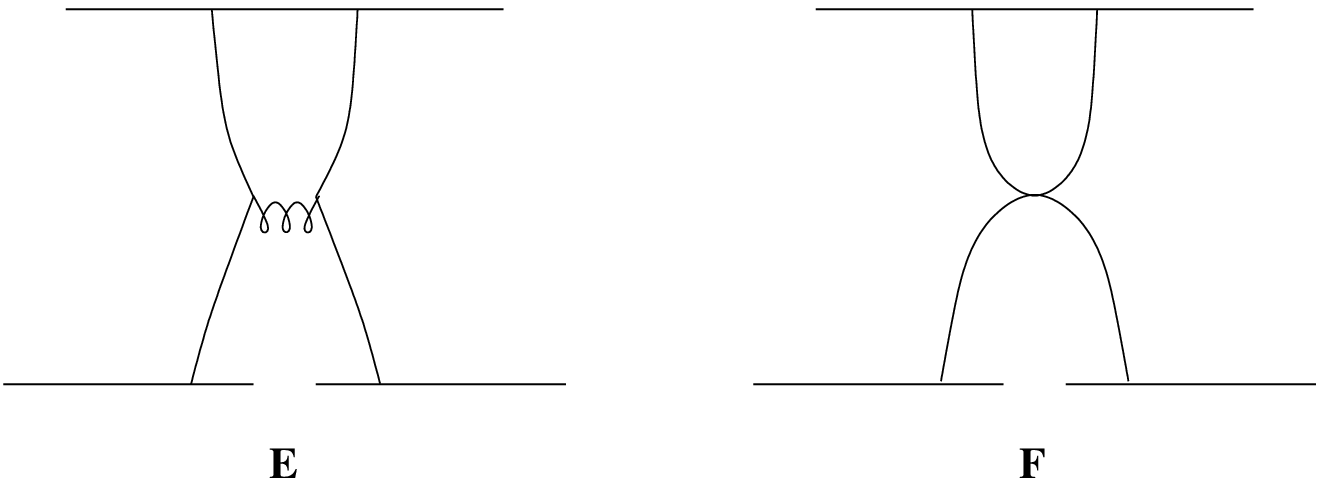}{3in} 
Another class of Feynman diagrams consists of those that act on
three operators.  These are shown in Fig.\ 2.  The corresponding action on
$\O[\psi_I]\O[\psi_J]\O[\psi_K]$ is given by
\eqn\threefeyn{\eqalign{{\rm Diagram\
E}~&=\;\de^{j_m}_{i_l}\de^{k_n}_{i_{l+1}}\,E(\x;\xx,\xxx)\,N^{-1}\,G(\x,\xx)\,G(\x,\xxx)\cr 
{\rm Diagram\ F}~&=\left (2\de^{k_{n}}_{i_l}\de^{j_m}_
{i_{l+1}} - \de^{j_m}_{i_l}\de^{k_n}_{i_{l+1}} - \de_{i_l
i_{l+1}}\de^{j_mk_n} \right )\,
F(\x;\xx,\xxx)\,N^{-1}\,G(\x,\xx)\,G(\x,\xxx)}}
Again, we can utilize the non-renormalization theorem for three-point
BPS operators \leemrs\ to constrain $E(\x;\xx,\xxx)$ and
$F(\x;\xx,\xxx)$ \dhofs.  The non-renormalization of the
three-point correlator $\bra \tr\,\phi^1\phi^2(\x)\,
\tr\,\phi^3\phi^1(\xx)\, \tr\,\phi^2\phi^3(\xxx) \ket$ implies the
following relation,
\eqn\threenonren{\eqalign{A'(\x,\xx)+&\,A'(\xx,\xxx)+A'(\x,\xxx)\, + \cr +& \,2\, \Big(
E(\x;\xx,\xxx) + E(\xx;\xxx,\x) + E(\xxx;\x,\xx)\, + \cr & \,\quad\quad +
F(\x;\xx,\xxx) + F(\xx;\xxx,\x) + F(\xxx;\x,\xx)\Big) = 0~~.}}

\subsec{An Example: Konishi Operator}

As a simple example, we work out explicitly the two- and three-point
function of the Konishi operator, $K=\frac{1}{\lam}\,\tr\,
\phi^i\phi^i\,$.  The free two- and three-point diagrams of the Konishi
operator are shown in Fig.\ 3.
\fig{Free Feynman Diagrams of the Konishi Operator.}{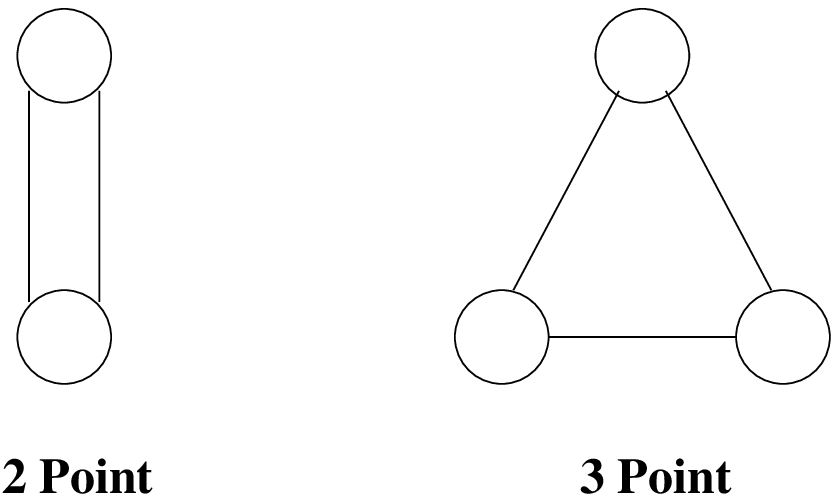}{2.0in}
At the free theory level, the two-point function is given by 
\eqn\ktwofree{\bra K(\x)K(\xx)\ket_{\rm free} 
= \frac{12}{(8\pi^2|x_{12}|^2)^2}~. }
The $O(\lam)$ correction gives
\eqn\ktwolam{\eqalign{\bra K(\x)K(\xx)\ket\Big|_{\lam} &=
\frac{12}{(8\pi^2|x_{12}|^2)^2}\left[2(A'+A-5B)\right] \cr
&=\frac{12}{(8\pi^2|x_{12}|^2)^2}\left[2(-6B)\right] \cr &=
\frac{12}{(8\pi^2|x_{12}|^2)^2}\left\{-12\lam\frac{1}{16\pi^2}\left[\ln\,|x_{12}\Lam|^2
- 1\right]\ri\}} }
where in the second line we have used the non-renormalization relation
\twononren\ and in the third line we have substituted in the expression
$B(\x,\xx) = \frac{\lam}{16\pi^2}\left[\ln|{x_{12}\Lam}|^2-1\right]$ (worked out in Appendix A using differential
regularization).  The Konishi operator perturbatively does not mix but
has a non-zero anomalous dimension.  Comparing \ktwofree\ and
\ktwolam\ with \twobare , we find that the normalization
$N_K=12/(8\pi^2)^2$, $a_K=\frac{3}{8\pi^2}$, and the anomalous dimension
$\gamma_K=\frac{3}{4\pi^2}\lam$.  For the
three-point function, we have
\eqn\kthreefree{\bra K(\x)K(\xx)K(\xxx)\ket_{\rm free} =
\frac{48}{N\,(8\pi^2)^3|x_{12}x_{13}x_{23}|^2}~,}
\eqn\kthreelam{\eqalign{
\bra K(\x)K(\xx)K(\xxx)\ket\Big|_{\lam} &=
\frac{48}{N\,(8\pi^2)^3|x_{12}x_{13}x_{23}|^2}\Big\{A'(\x,\xx)+\,A'(\xx,\xxx)+A'(\x,\xxx)
+\cr  &\quad\quad + 2\, [
E(\x;\xx,\xxx) + E(\xx;\xxx,\x) + E(\xx;\x,\xx)]+ \cr &\quad\quad + 2\,(2-1-6) [F(\x;\xx,\xxx)+
F(\xx;\xxx,\x) + F(\xxx;\x,\xx)]\Big\} \cr
&=\frac{48}{N\,(8\pi^2)^3|x_{12}x_{13}x_{23}|^2}\Big\{-12[F(\x;\xx,\xxx)+
F(\xx;\xxx,\x) + \cr &\qquad\qquad\qquad\qquad\qquad\qquad +
F(\xxx;\x,\xx)]\,\Big\}\cr
&=\frac{48}{N\,(8\pi^2)^3|x_{12}x_{13}x_{23}|^2}\big(-\frac{3}{4\pi^2}\ln\,|x_{12}x_{13}x_{23}
\Lam^3|\, \big) } }
where again, we have used the non-renormalization relation
\threenonren\ and the explicit expression
$F(\xxx;\x,\xx)=\frac{\lam}{32\pi^2} \ln\lf|{x_{13}x_{23}\La\o x_{12}}\ri|^2
$ (from Appendix A).  Comparing \kthreelam\ with \threebare, we obtain again
$\gamma_K=\frac{3}{4\pi^2}\lam$ and $C=0$, implying
$c^1_{KKK}=-3a_K=-\frac{9}{8\pi^2}$.  These values agree with the
results for the Konishi operator in \biakrs.  

The above Konishi calculation highlights two important features for
calculating the one-loop correction, $c^1_{\al\bt\rho}$.
First, notice that we were able to express any dependence on functions
$A'$, $A$, and $E$ in terms of $B$ and $F$ using the
non-renormalization relations \twononren\ and \threenonren.  Indeed,
in general, only $B$ and $F$ will appear in any two- or three-point
function calculation involving scalar operators.  Any contribution from $A'$, 
$A$, and $E$, which contains gauge boson exchange, can always be
rewritten in terms of $B$ and $F$.  As argued in \ryzhov, this must be
the case because the functions $A'$, $A$, and $E$ contains
terms that are dependent on the gauge fixing parameter.  Although the
gauge boson exchange Feynman diagrams may have a contribution to the
one-loop correction, the functions themselves can not contribute
directly to any gauge independent two- or three-point function.

Secondly, we emphasize that though the value of $a_K$ is scheme
dependent, $c^1_{KKK}$ is scheme independent.  For if instead of using
an explicit expression for $B$ and $F$, we consider the scheme
independent form $B(\x,\xx)= b_0 +
\frac{\lam}{16\pi^2}\ln\,|x_{12}\Lam|^2$ and $F(\xxx;\x,\xx)= f_0 +
\frac{\lam}{32\pi^2} \ln\,\left|\frac{x_{13}x_{23}\Lam}{x_{12}}\right|^2$,
leaving the constant $b_0$ and $f_0$ arbitrary.  This leads to
$c^1_{KKK}=-18(2\,f_0 - b_0)$.  But as pointed out in \dhoryz, the expression
\eqn\schemeindep{F(\xxx;\x,\xx) + F(\x;\xxx,\xx) - B(\x,\xxx) = 2f_0 -
b_0} does not depend on the regulator and therefore is a scheme independent
quantity.  The precise constant can be found, for example, using differential
regularization (as in Appendix A) with the result $2f_0 -
b_0=\frac{\lam}{16\pi^2}$.  Thus, in calculating $c^1_{\al\bt\rho}$ we
have the freedom of working in any scheme.  In particular, we can work in the
scheme where $f_0=0$, as in the Konishi example above, and in fact not
consider any three-point Feynman diagrams contributions.  As well, we may
choose $b_0=0$ and not consider any two-point Feynman diagram
contributions in calculating $c^1_{\al\bt\rho}$.  In the next
subsection, we will show by applying conformal invariance that
$c^1_{\al\bt\rho}$ is always proportional to $2f_0-b_0$.

\subsec{Formulas for $c^1_{\al\bt\rho}$}

That the functional dependence of the first order corrections to two-
and three-point functions can only take on the form $B$ and $F$ is a powerful
supersymmetry constraint.  Combining this with constraints from
conformal symmetry gives general formulas for $c^1_{\al\bt\rho}$ up to
operator dependent combinatorial factors.  We start with the two
point function.  The one-loop correction comes only from the $B$ function
and implies the general form 
\eqn\twogen{\eqalign{\bra\bO_\al(\x)\O_\al(\xx)\ket
&=\frac{N_\al^2}{|x_{12}|^{2\De_{0\al}}} \left[1+ b_\al\, B(\x,\xx)\,\right]\cr
&=\frac{N_\al^2}{|x_{12}|^{2\De_{0\al}}} \left[1+ b_\al\, b_0 
+ \frac{b_\al\lam}{16
\pi^2}\ln\,\left|x_{12}\Lam\right|^2\,\right] } }
where $b_\al$ is an operator dependent combinatorial constant and we have
substituted in $B(x_1,x_2)=b_0+{\la\o16\pi^2}\ln|x_{12}\Lam|^2$.
Comparing \twogen\ with the required form in \twobare\ gives
\eqn\twoformula{a_\al={b_0 \o 2}\,b_\al~, \qquad
\gamma_\al=-\frac{\lam}{16\pi^2}\, b_\al ~,}
This gives the following relationship between $\ga_\al$ and $a_\al$, 
\eqn\agam{\ga_\al=-\frac{\lam}{16\pi^2}\left(\frac{2a_\al}{b_0}\right)=-\frac{\lam}{16\pi^2}\,\ta_\al~,}
where we have defined $\ta_\al=2a_\al/{b_0}$.  From \twoformula,
calculating the anomalous dimension is just computing the constant
$b_\al$.  An explicit form for this constant in
the planar limit will be given in the next section.  

Now we proceed with the three-point function calculation.  In
addition to two-point $B$ functional dependences, we have additional
three-point $F$ functional dependences.  The first order
three-point function can be simply expressed as 
\eqn\threeptc{\eqalign{\bra
\bO_\al(\x)\O_\bt(\xx)\O_\rho(\xxx)\ket& =
\frac{C^0}{|x_{12}|^{\De_{0\al}+\De_{0\bt}-\De_{0\rho}}|x_{13}|^{\De_{0\al}+\De_{0\rho}-\De_{0\bt}}|x_{23}|^{\De_{0\bt}+\De_{0\rho}-\De_{0\al}}}\times\cr\times\Big\{
1 +& \big[\b B(\x,\xx) +\bb B(\xx,\xxx) +\bbb B(\xxx,\x)\big] \cr +& \big[\f  F(\x;\xx,\xxx) + \ff F(\xx;\xxx,\x) +\fff F(\xxx;\x,\xx)\big] \Big\} }} 
where $f^i_{jk}$ and $b_{ij}$ are again combinatorial constants
dependent on the three operators.  These constants specify the strength of
the associated Feynman diagram contributions.  For example, $\b$
is associated with two-point Feynman diagrams that contracts between
$\bO_\al$ and $\O_\bt$, and $\f$ is associated with three-point
Feynman diagram that have two contractions with $\bO_\al$ and one each with
$\O_\bt$ and $\O_\rho$.  Substituting the expressions for
$B(x_1,x_2)=b_0+{\la\o16\pi^2}\ln|x_{12}\Lam|^2$ and
$F(x_3;x_1,x_2)=f_0+{\la\o32\pi^2}\ln\lf|{x_{31}x_{23}\Lam\o
x_{12}}\ri|^2$ into \threeptc, we obtain 
\eqn\threeptcc{\eqalign{\bra
\bO_\al(\x)\O_\bt(\xx)&\O_\rho(\xxx)\ket =
\frac{C^0}{|x_{12}|^{\De_{0\al}+\De_{0\bt}-\De_{0\rho}}|x_{13}|^{\De_{0\al}+\De_{0\rho}-\De_{0\bt}}|x_{23}|^{\De_{0\bt}+\De_{0\rho}-\De_{0\al}}}\times\cr\times\Big\{
&1 + (\b+\bb+\bbb)b_0 +(\f+\ff+\fff)f_0  \cr
&~+\frac{\lam}{16\pi^2}\big[2\,\b\,\ln|x_{12}\Lam| +
2\,\bb\,\ln|x_{23}\Lam|+ 2\,\bbb\,\ln|x_{31}\Lam| \cr&
~~\quad\quad\quad\quad + \f\,{\ln|\frac{x_{12}x_{31}\Lam}{x_{23}}|}
+\ff\,{\ln|\frac{x_{23}x_{12}\Lam}{x_{31}}|}+\fff\,{\ln|\frac{x_{31}x_{23}\Lam}{x_{12}}|}~\big]~\Big\}~. } }
We can compare \threeptcc\ with the expected form for the three
point function in \threebare.  Together with \agam, we arrive at the
following relations
\eqn\garel{\eqalign{\ga_\al&=-\frac{\lam}{16\pi^2} \left[\f + \b +
\bbb\right] = -\frac{\lam}{16\pi^2} \ta_\al \cr
\ga_\bt&=-\frac{\lam}{16\pi^2} \left[\ff + \b + \bb\right] =
-\frac{\lam}{16\pi^2} \ta_\bt\cr 
\ga_\rho&=-\frac{\lam}{16\pi^2} \left[\fff + \bbb + \bb \right] = 
-\frac{\lam}{16\pi^2} \ta_\rho~.
}}
The above relations for eigen-operators are direct consequences of
conformal invariance and supersymmetry.  We can rewrite \garel\ as a
relationship between two- and three-point constants as follows 
\eqn\aeq{\f=\ta_\al - \b - \bbb ,\quad  \ff=\ta_\bt - \b -
\bb ,\quad \fff=\ta_\rho - \bbb - \bb ~.
}  
\foot{Similar equations to \aeq\ but with all $\ta$'s set to zero
can be found in \dhoryz.}  Moreover, from \threeptcc,
\eqn\Cexp{\lam\, C^1=\lam\,(c^1_{\al\bt\rho}+a_\al+a_\bt+a_\rho) =
(\b+\bb+\bbb)b_0 + (\f+\ff+\fff)f_0~.}
This implies with \garel,
\eqn\cexp{\eqalign{\lam\,c^1_{\al\bt\rho}&=
\lam\left(C^1-a_\al-a_\bt-a_\rho\right)\cr &=
b_0(\b+\bb+\bbb)+f_0(\f+\ff+\fff)\cr&\qquad\qquad -
\frac{b_0}{2}\left(\f+\ff+\fff+2 (\b+\bb+\bbb) \right)\cr
&=\frac{2f_0 - b_0}{2}\left(\f+\ff+\fff\right).}}
Therefore, we have obtained the general expression for
the first order correction to the three-point coefficient and have found
that it is indeed proportional to the scheme independent expression,
$2f_0 - b_0$.  If we now substitute the calculated value of
$2f_0 - b_0=\frac{\lam}{16\pi^2}$ into \cexp, we arrive at our main result  
\eqn\coneinf{
\lam\,c^1_{\al\bt\rho}= \frac{\lam}{32\pi^2}\left(\f+\ff+\fff\right)~,
}
which can be equivalently expressed using \aeq\ in the form
\eqn\ctwoexp{\la\, c^1_{\al\bt\rho}=-\hf(\ga_\al+\ga_\bt+\ga_\rho)
-{\la\o16\pi^2}(\b+\bb+\bbb)~.}

We have obtained \coneinf\ and \ctwoexp\ without setting either $f_0$
or $b_0$ equal to zero.  Hence, in effect, we have summed up all possible
Feynman diagrams.  But clearly from \coneinf, we could have worked in
the scheme $(b_0,f_0)=(0,\frac{\lam}{32\pi^2})$ and just summed up
three-point Feynman diagrams.  Similarly,
summing up only two-point Feynman diagrams in the scheme
$(b_0,f_0)=(-\frac{\lam}{16\pi^2},0)$ will give the identical result for
$c^1_{\al\bt\rho}$ as expressed in \ctwoexp.  Thus, these two
schemes present two very different prescriptions for calculating
$c^1_{\al\bt\rho}$.

The formula for $c^1_{\al\bt\rho}$ simplify further for the extremal
three-point function.  Suppose $\De_{0\al}>\De_{0\bt},\De_{0\rho}$, then
the three-point function is called extremal if
$\De_{0\al}=\De_{0\bt}+\De_{0\rho}$.  In this special case, there are
no contractions between operators $\O_\bt(\xx)$ and $\O_\rho(\xxx)$ at
the free level.  This lack of contractions constrains the types
of contributing Feynman diagrams at $O(\lam)$ and results in
$\ff=\fff=\bb=0$.  The three constraint relations in \aeq\ then becomes simply
\eqn\aexeq{\f=\ta_\al-\b-\bbb~,\qquad\ta_\bt=\b~,\qquad \ta_\rho=\bbb~.}
This implies
\eqn\exrel{\f=\ta_\al-\ta_\bt-\ta_\rho=-\frac{16\pi^2}{\lam}(\ga_\al-\ga_\bt-\ga_\rho)} 
where in the last equality, we have used \agam.   Substituting \exrel\ into
\cexp, we obtain the very simple expression 
\eqn\cexexp{\lam\,c^1_{\al\bt\rho}=\frac{\lam}{32\pi^2}\,\f={1\o 2}(\ga_\bt+\ga_\rho-\ga_\al)~.}
Thus, we have shown that the first order correction to the extremal
three-point coefficient depends only on the anomalous dimensions of
the three scalar primary operators.  This result is valid at
finite $N$ and, in particular, applies for operators in the $SU(2)$
subsector.

\newsec{$c^1_{\al\bt\rho}$ in the Planar Limit}

In the previous section, using constraints from superconformal invariance,
we wrote down the general form of $c^1_{\al\bt\rho}$ in terms of
anomalous dimensions and the combinatorial constants, $b_{ij}$ and
$f^{i}_{jk}$.  In general, these constants are not easy to calculate
at finite $N$ in the $SU(2)$ subsector.  However, at the planar limit,
it is possible to write down explicit expressions for these
constants in the $SO(6)$ subsector.  In this section, we will take the
planar limit and consider single trace scalar operators only.   We
will calculate explicitly $c^1_{123}$ for three generic operators,
$\bO_1[\psi_I](\x)$, $\O_2[\psi_J](\xx)$, and $\O_3[\psi_K](\xxx)$,
with length $L_1$, $L_2$, and $L_3$, respectively.  Below, we set
up our notations by first considering two-point functions and free three-point
functions.  We then proceed to discuss $c^1_{123}$ for the
non-extremal case, $L_1>L_2+L_3$, followed by the extremal case,
$L_1=L_2+L_3$.

\subsec{Two-Point Functions and Free Three-Point Functions}

We write each operator in the form
\eqn\opdeff{\eqalign{\O[\psi_I]&=\frac{1}{\lam^{L/2}}\,\psi_{i_1\ldots i_L}
\tr\,\phi^{i_1}\cdots\phi^{i_L}\cr &=\frac{1}{\lam^{L/2}
L}\tpsi_{i_1\ldots i_L} \tr\,\phi^{i_1}\cdots\phi^{i_L} ~,}}
where in the second line we have introduced a normalized coefficient
$\tpsi= L \psi$.  We emphasize that all indices in $I=\{i_1,\ldots,i_L\}$ are
summed over in \opdeff, and throughout, we will not explicitly write out
the summation symbol.  Also, the presence of the trace
implies that $\psi_{i_1\ldots i_L}$ must be invariant under cyclic
permutation of the indices.  Thus, for example,
\eqn\opex{\eqalign{\tr\, \phi^1\phi^2\phi^3& =
\psi_{i_1i_2i_3}\,\tr\,\phi^{i_1}\phi^{i_2}\phi^{i_3}\cr &=\psi_{123}\,
\tr\, \phi^1\phi^2\phi^3 + \psi_{231}\, \tr\,\phi^2\phi^3\phi^1 +
\psi_{312}\, \tr\,\phi^3\phi^1\phi^2 \cr &=\frac{1}{3}\left(\tpsi_{123}\,
\tr\, \phi^1\phi^2\phi^3 + \tpsi_{231}\, \tr\,\phi^2\phi^3\phi^1 +
\tpsi_{312}\, \tr\,\phi^3\phi^1\phi^2\right) }}
with $\psi_{123}=\psi_{231}=\psi_{312}=1/3$ and
$\tpsi_{123}=\tpsi_{231}=\tpsi_{312}=1$.  In writing out the form of
the two and three-point functions, it is useful to establish
conventions for labeling the indices of the $\psi$ coefficients as
shown in Fig.\ 4.  The two-point function at one-loop can then be expressed as
\eqn\twoptc{\eqalign{\bra\bO_\al[\psi_{I'}](\x)\O_\al[\psi_{I}](\xx)\ket&=
\bra~\frac{1}{\lam^{L/2}}\psi_{i'_L\ldots i'_1}
\tr\,\phi^{i'_L}\cdots\phi^{i'_1}(\x)\frac{1}{\lam^{L/2}}\psi_{i_1\ldots i_L}
\tr\,\phi^{i_1}\cdots\phi^{i_L}(\xx)\ket\cr&=
\frac{\tpsi_{I'}\psi_{I} \I_2 + \tpsi_{I'}\tpsi_{I}
(2\PP-2\I-\K)^{i'_1i'_2}_{i_1i_2}\I'_2 B(\x,\xx)}{[8\pi^2|x_{12}|^2]^L}} } 
where
\eqn\opdef{\PP_{ij}^{kl} = \de_{i}^{l}\de_{j}^{k}~,\quad \I_{ij}^{kl}
= \de_{i}^{k}\de_{j}^{l}~,\quad \K_{ij}^{kl} =
\de_{ij}\,\de^{kl}~.} 
\eqn\ideftwo{\I_2=\de^{i'_1}_{i_1}\de^{i'_2}_{i_2}\ldots\de^{i'_L}_{i_L},
\quad \I'_2=\de^{i'_3}_{i_3}\ldots\de^{i'_L}_{i_L}~.}
The ordering of the indices in line one of \twoptc\ follows from Fig.\
4, using a ``right-handed'' counter-clockwise convention.  The
ordering for $\psi_{I'}$ is reversed to take account of hermitian
conjugation.  In the second line of \twoptc, we have summed up all
contributing two-point Feynman diagrams.  $\PP$, $\I$, and $\K$ are the
permutation, identity, and trace operators typical of nearest neighbor
spin chain interactions.  $\I_2$ represents full free contractions of
the indices of the two operators and the primed $\I'_2$, written
explicitly in \ideftwo, represents free contractions of all indices
that have not yet been contracted.  Substituting
$B(x_1,x_2)=b_0+{\la\o16\pi^2}\ln|x_{12}\Lam|^2$ into \twoptc\ and
comparing with the expected form for two-point functions in \twogen,
we find for the anomalous dimension \agam,
\eqn\gacalc{\ga_\al=-\frac{\lam}{16\pi^2}
\frac{\tpsi_{I'}\tpsi_{I}(2\PP-2\I-\K)^{i'_1i'_2}_{i_1i_2}\I'_2}{\tpsi_{I'}\psi_{I}\I}
= -\frac{\lam}{16\pi^2} \ta_\al ~.}
Recalling that $\O_\al[\psi_I]$ is an eigen-operator and also that the
indices of $\psi$ are cyclically invariant, \gacalc\ is indeed
identical to the anomalous dimension formula given in \minzar.
\fig{Labeling of the Free Two- and Three-Point
Operators.}{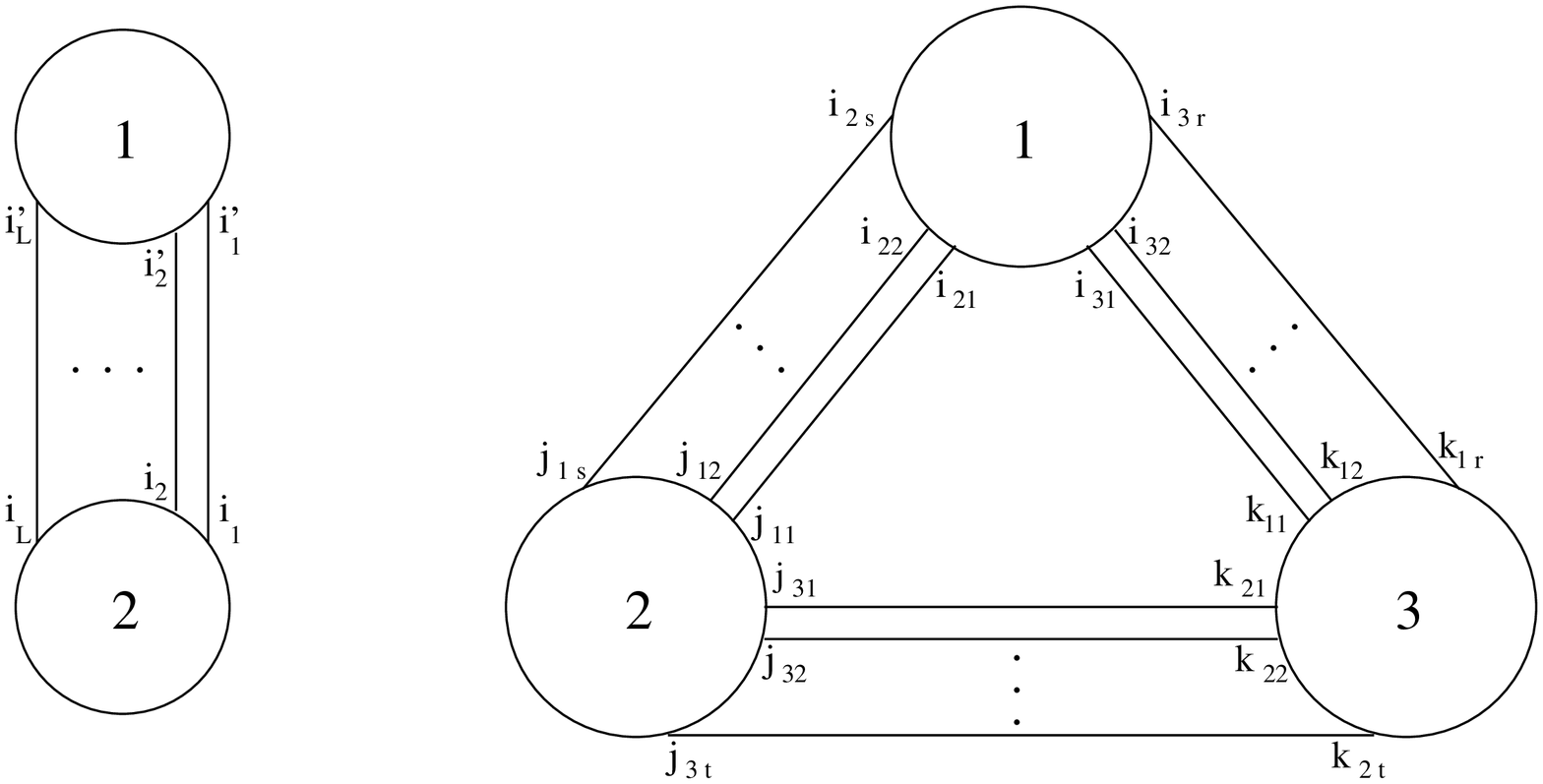}{4in} 

For the three-point function, the $\psi$ indices are labeled as follows,
\eqn\notathree{\eqalign{\bO_1[\psi_I]:&\qquad \psi_I =
\psi_{i_{31}i_{32}\ldots\, i_{3r}i_{2s}\ldots\, i_{22} i_{21}}\cr
\O_2[\psi_J]:&\qquad \psi_J =\psi_{j_{11}j_{12}\ldots\,
j_{1s}j_{3t}\ldots\, j_{32}j_{31}}\cr \O_3[\psi_K]:&\qquad \psi_K
=\psi_{k_{21}k_{22}\ldots\,k_{2t}k_{1r}\ldots\,k_{12}k_{11}}}}
where for example $i_{32}$ signifies the ``2nd'' contraction starting
from $I$ to the ``3rd'' operator.   Note that $r$, $s$, and $t$ are the number of contractions between operators 3-1, 1-2, and 2-3 
respectively.  These three numbers are uniquely determined by the
lengths of the operators by the following relations
\eqn\rst{r=\frac{1}{2}\left(L_3 + L_1 -
L_2\right)\,,\quad s=\frac{1}{2}\left(L_1 + L_2 -
L_3\right)\,,\quad t=\frac{1}{2}\left(L_2 + L_3 -
L_1\right)~.} 
Moreover, if $r$, $s$, and $t$ as defined in \rst\ are not all
integers, then the three-point function must be zero.  For at the free
level, the full contraction is given by
\eqn\contracttt{\I_3=\de_{i_{31}}^{k_{11}}\de_{i_{32}}^{k_{12}}\cdots\,\de_{i_{3r}}^{k_{1r}}
\,\de_{i_{21}}^{j_{11}}\de_{i_{22}}^{j_{12}}\cdots\,\de_{i_{2s}}^{j_{1s}}
\,\de_{j_{31}}^{k_{21}}\de_{j_{32}}^{k_{22}}\cdots\,\de_{j_{3t}}^{k_{2t}}~.}
We can use $\I_3$ to simply express the free three-point function as
\eqn\freethree{\bra
\bO_1[\psi_I](\x)\O_2[\psi_J](\xx)\O_3[\psi_K](\xxx)\ket_{\rm free} =
\frac{\tpsi_I\tpsi_J\tpsi_K\I_3}{N\,(8\pi^2)^{r+s+t}|x_{31}|^{2r}|x_{12}|^{2s}|x_{23}|^{2t}}~.} 
And as a contribution to the operator production expansion (OPE),
\freethree\ implies
\eqn\freeope{\O_2[\psi_J](\xx)\O_3[\psi_k](\xxx) \sim
\frac{\tpsi_J\tpsi_K\de_{j_{31}}^{k_{21}}\de_{j_{32}}^{k_{22}}\cdots\,\de_{j_{3t}}^{k_{2t}}}{N(8\pi^2)^t|x_{23}|^{2t}}\tr\;\phi^{j_{11}}\phi^{j_{12}}
\cdots\phi^{j_{1s}}\phi^{k_{1r}}\cdots\phi^{k_{12}}\phi^{k_{1r}}(\xxx)
+ \cdots }
with the value of $t$ fixing the length of the resulting operator.

\subsec{Non-extremal Three-Point Function}

The expression for $c^1_{123}$ can be expressed explicitly in the planar
limit.  Contribution to $c^1_{123}$ consists of three
families of two-point Feynman diagrams and six different E and F types
three-point Feynman diagrams.  Some of which are
shown in Fig.\ 5.

\fig{Examples of planar diagrams contributing to $c^1_{123}$.  (a) A
two-point B type diagram. (b) Two three-point F type
diagrams.}{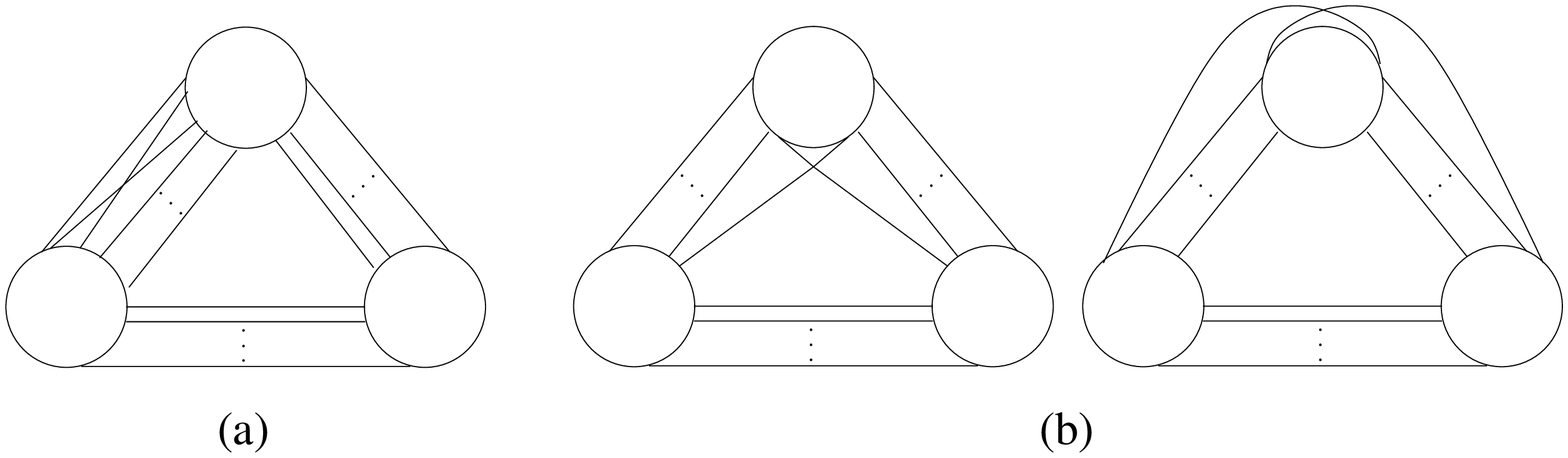}{4.5in} 

Again, the nonzero contribution to $c^1_{123}$ consists only of
two-point interaction with $B$ functional dependence and three-point
interaction with $F$ functional dependence.  The one-loop order
three-point function is thus expressed as 
\eqn\threeptc{\eqalign{\bra
\bO_1[\psi_I](\x)\O_2[\psi_J](\xx)&\O_3[\psi_K](\xxx)\ket =
\frac{\tC^0}{N\,(8\pi^2)^{r+s+t}|x_{31}|^{2r}|x_{12}|^{2s}|x_{23}|^{2t}}\times\cr\times\Big\{
1 +& \big[\b B(\x,\xx) +\bb B(\xx,\xxx) +\bbb B(\xxx,\x)\big] \cr +&
\big[\f  F(\x;\xx,\xxx) + \ff F(\xx;\xxx,\x) +\fff F(\xxx;\x,\xx)\big]
\Big\}} }
where we have defined $\tC^0=C^0 N(8\pi^2)^{r+s+t}=\tpsi_I\tpsi_J\tpsi_K\I_3$.
Summing up all contributions, we find the following
expressions for the constants
\eqn\fbdef{\eqalign{\b&=\sum_{l=1}^{s-1}\,(2\PP-2\I-\K)_{i_{2\,l}i_{2\,l+1}}^{j_{1\,l}j_{1\,l+1}}\frac{\tpsi_I\tpsi_J\tpsi_K \I'_3}{\tC^0}\cr
\bb&=\sum_{l=1}^{t-1}\,(2\PP-2\I-\K)_{j_{3\,l}j_{3\,l+1}}^{k_{2\,l}k_{2\,l+1}}\frac{\tpsi_I\tpsi_J\tpsi_K \I'_3}{\tC^0}\cr
\bbb&=\sum_{l=1}^{r-1}\,(2\PP-2\I-\K)^{i_{3\,l}i_{3\,l+1}}_{k_{1\,l}k_{1\,l+1}}\frac{\tpsi_I\tpsi_J\tpsi_K\I'_3}{\tC^0}\cr
\f&=\left[(2\PP-2\I-\K)_{i_{21}i_{31}}^{j_{11}k_{11}}\I'_3+(2\PP-2\I-\K)_{i_{3r}i_{2s}}^{k_{1r}j_{1s}}\I'_3\right] \frac{\tpsi_I\tpsi_J\tpsi_K}{\tC^0}\cr
\ff&=\left[(2\PP-2\I-\K)_{j_{31}j_{11}}^{k_{21}i_{21}}\I'_3+(2\PP-2\I-\K)_{j_{1s}j_{3t}}^{i_{2s}k_{2t}}\I'_3\right]
 \frac{\tpsi_I\tpsi_J\tpsi_K}{\tC^0}\cr
\fff&=\left[(2\PP-2\I-\K)_{k_{11}k_{21}}^{i_{31}j_{31}}\I'_3+(2\PP-2\I-\K)_{k_{2t}k_{1r}}^{j_{3t}i_{3r}}\I'_3\right]
\frac{\tpsi_I\tpsi_J\tpsi_K}{\tC^0}~~.}}
In above, ${\cal I}'_3$ is defined to be the required free
contractions that fully contract the remaining indices and $\I_3$ is
the full free contraction defined in \contracttt. Thus, for example,
with all the contractions explicitly written out,
\eqn\Idef{\eqalign{\f=\frac{\tpsi_I\tpsi_J\tpsi_K}{\tC^0}&
\left[(2\PP-2\I-\K)_{i_{21}i_{31}}^{j_{11}k_{11}}\I_{i_{3r}i_{2s}}^{k_{1r}j_{1s}}+(2\PP-2\I-\K)_{i_{3r}i_{2s}}^{k_{1r}j_{1s}}\I_{i_{21}i_{31}}^{j_{11}k_{11}}\right]\cr&\times\de_{i_{22}}^{j_{12}}\de_{i_{23}}^{j_{13}}\cdots\,\de_{i_{2\,s-1}}^{j_{1\,s-1}}
\,\de_{i_{32}}^{k_{12}}\de_{i_{33}}^{k_{13}}\cdots\,\de_{i_{3\,r-1}}^{k_{1\,r-1}}
\,\de_{j_{31}}^{k_{21}}\de_{j_{32}}^{k_{22}}\cdots\,\de_{j_{3\,t}}^{k_{2\,t}} }}
where $\I_{ij}^{kl}= \de_{i}^{k}\de_{j}^{l}$.  The constants in \fbdef\
together with either \coneinf\ or \ctwoexp\ gives the explicit formula
for $c^1_{123}$.  In the next section, we will interpret the planar
expressions from the spin chain perspective.

\subsec{Extremal Three-Point Function}

\fig{Non-nearest neighbor diagram for extremal three-point
functions.}{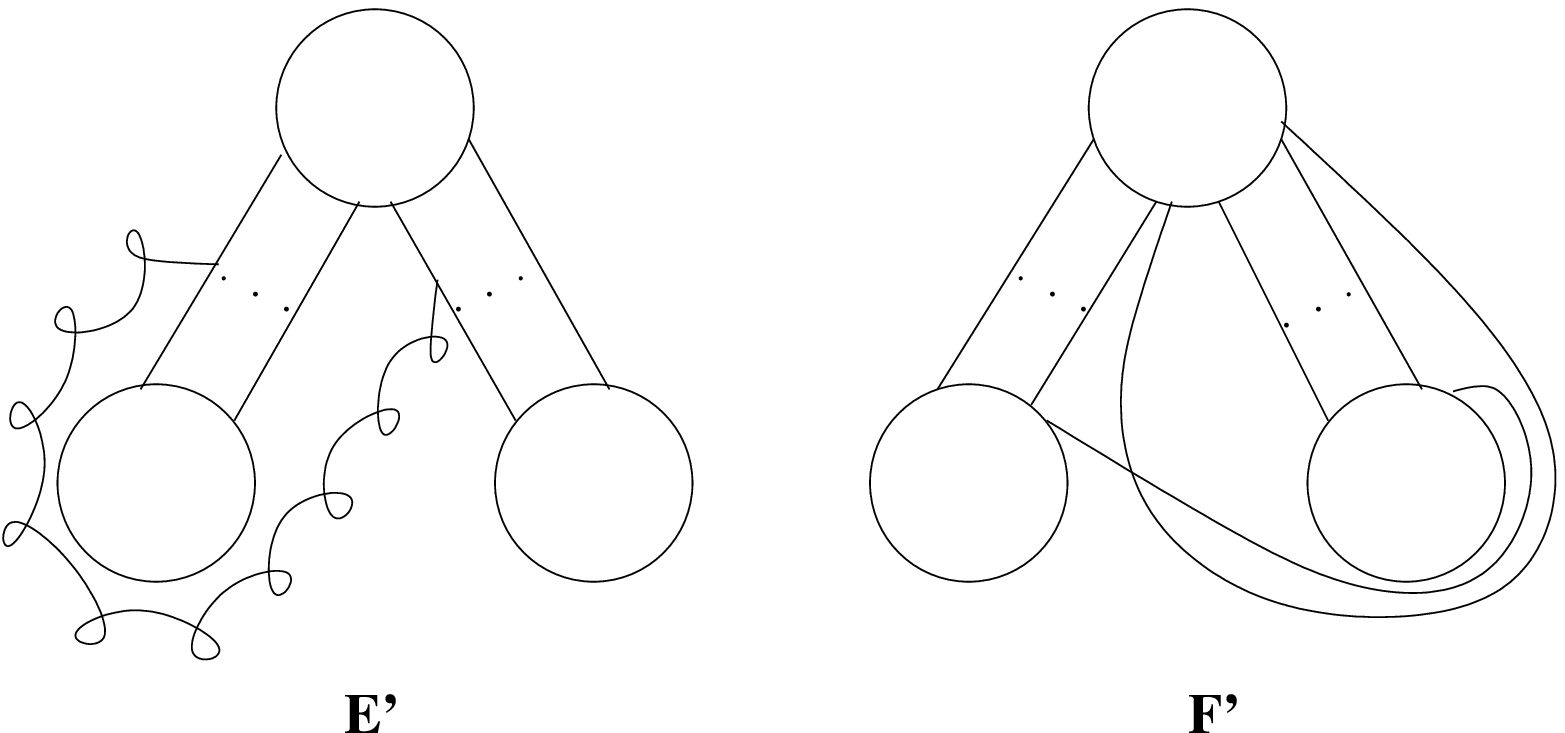}{3in}
For the extremal three-point function, with $L_1=L_2+L_3$, the number
of contractions between operators 2 and 3 is zero (i.e. $t=0$).  We
have found that $c^1_{123}$ is simply given by \cexexp\ in terms of
the anomalous dimensions of the three scalar primary operators.
Nevertheless, it is interesting to write down the explicit expressions
for the three nonzero constants - $\f$, $\b$, and $\bbb$ - in the
planar limit.  Interestingly, they are not
those in \fbdef.  For in the extremal case, two subtleties must be
incorporated into the calculation.  First, at the free
level, the mixing of $\O_1[\psi_I](\x)$ with the double trace
operator $\O_1'(\x)=\O_2[\psi_J]\O_3[\psi_K](\x)$ must be taken
into account \dhofmmr.  That the mixing coefficient contains
a factor of $N^{-1}$ is compensated by the fact that the free
three-point function of $\O_1'(\x)$ with $\O_2(\xx)$ and $\O_3(\xxx)$
is $O(N^0)$ instead of $O(N^{-1})$.  Though $\O_1'$ makes no contribution
to first order correction $C^1$, $C^0$ can not longer be expressed
simply as in \freethree.  Second, in addition to the Feynman diagrams
in Fig.\ 5, there are planar non-nearest neighbor diagrams that also
must be considered.  The new diagrams, E' and F', are shown in Fig.\
6.  The contributions of diagrams E' exactly cancel out those of
diagrams E of Fig.\ 2.  However, diagram F' gives
a nonzero non-nearest neighbor contribution not present in the
non-extremal case.  Adding up the contributing Feynman diagrams, we find
\eqn\fbdefex{\eqalign{\f&=\left[(2\PP-\I-\K)_{i_{21}i_{31}}^{j_{11}k_{11}}\I_3'
+ (2\PP-\I-\K)_{i_{3r}i_{2s}}^{k_{1r}j_{1s}}\I_3'+(2\PP -
\I-\K)_{j_{11}i_{21}}^{i_{3r}k_{1r}}\I_3'\right]\frac{\tpsi_I\tpsi_J\tpsi_K}{\tC^0}\cr
\b&=\left[\sum_{l=1}^{s-1}\,(2\PP-2\I-\K)_{i_{2\,l}i_{2\,l+1}}^{j_{1\,l}j_{1\,l+1}}\I'_3
+ (2\PP-2\I-\K)_{i_{2\,s}i_{21}}^{j_{1\,s}j_{11}}\I'_3\right]\frac{\tpsi_I\tpsi_J\tpsi_K}{\tC^0}\cr
\bbb&=\left[\sum_{l=1}^{r-1}\,(2\PP-2\I-\K)^{i_{3\,l}i_{3\,l+1}}_{k_{1\,l}k_{1\,l+1}}\I'_3+(2\PP-2\I-\K)_{i_{3\,r}i_{31}}^{k_{1\,r}k_{11}}\I'_3\right]\frac{\tpsi_I\tpsi_J\tpsi_K}{\tC^0}}} 
where once again, $\I_3'$ denotes the required free contractions to
fully contract each term.  In $\f$, the third term corresponds to the
contribution from the non-nearest neighbor F' type Feynman diagrams in
Fig.\ 6.  The expressions in \fbdefex\ must satisfy \aexeq, which also
gives an indirect way of calculating $\tC^0$.

\newsec{Planar One-Loop Correction and Integrable Spin Chains}

The general formulas for $c^1_{123}$ in the previous section are
given in explicit forms in the planar limit.  In light of the connection of
operators with states in $SO(6)$ integrable spin chain in the planar
limit \minzar, we will interpret $c^1_{123}$ from the spin chain
perspective.

\subsec{Mapping SYM into Spin Chain}

As is clear from the one-loop planar formula in
\gacalc, the anomalous dimension is just a combinatorial factor
depending only on the $SO(6)$ chain of indices of the operator and
nothing else.  Therefore, focusing only on the $SO(6)$ indices, we can
naturally map $\O[\psi_I]$ to a spin chain state
\eqn\opspin{|\Psi\ket=\tpsi_{i_1\dots i_L} |i_1\ldots i_L\ket~}
where, as before, $\tilde{\psi}_{i_1\dots i_L}$ is invariant under cyclic
permutation.   Note that $|i_1\ldots i_L\ket$ effectively spans the Hilbert
space $\CH=V^{\tens L}$ where $V={\bf R}^6$.  We will define the conjugate of
$|\Psi\ket$ as \foot{In the $SO(6)$ subsector, eigenstates of the planar
anomalous dimension matrix can be taken to be real vectors, since that
matrix is real symmetric.  Therefore, we can impose that the coefficients
$\tilde{\psi}_{i_1\dots i_L}$ in \opspin\ are real numbers.} 
\eqn\opspinh{\bra\Psi|=\bra i_1\ldots i_L|\tpsi_{i_L\dots i_1}}
where we have reversed the order of the indices so that the inner
product is given by  $\bra a_1\ldots a_r|b_1\ldots b_r\ket=
\cob^{a_1}_{b_1}\ldots\cob^{a_r}_{b_r}$.  As pointed out by
Minahan and Zarembo \minzar, the planar one-loop anomalous dimension
matrix is just proportional to the hamiltonian of the $SO(6)$ 
integrable closed spin chain.  In our notations, the anomalous
dimension \gacalc\ can be written simply as
\eqn\gaH{
\ga=\frac{\lam}{16\pi^2}\frac{\bra\Psi|H|\Psi\ket}{\bra\Psi|\Psi\ket}}
where $|\Psi\ket$ is assumed to be an eigenstate of the $SO(6)$
integrable closed hamiltonian,
\eqn\ham{H = \sum_{l=1}^L
\left(\K_{l,l+1}+2\I_{l,l+1}-2\PP_{l,l+1}\right) ~.}
Being integrable, the eigenvalues and eigenstates of $H$ can be found
using the algebraic Bethe ansatz techniques.

Now, for the one-loop correction $c^1_{123}$, the explicit planar
expressions for both non-extremal and extremal cases also only depend on
the $SO(6)$ indices of the three eigen-operators. (See eqs. \coneinf,
\cexexp, and \fbdef.)  This again implies that there is an $SO(6)$
spin chain interpretation.  For the extremal case, the integrable
structure is self-evident since, as in \cexexp,  $c^1_{123}$ is just a
linear combination of anomalous dimensions.  For the non-extremal
case, we will show the integrability in the spin chain description
using the two-point interaction prescription of \ctwoexp. 

In the explicit expressions for $\b, \bb$, and $\bbb$ in \fbdef, the action on
the $SO(6)$ indices is similar to \ham\ except that there is no
interaction between sites at the two ends.  For example, for $\bbb$,
sites $k_{1r}$ and $k_{11}$ do not interact with each other.  A
hamiltonian such as 
\eqn\Hs{H_r=\sum_{l=1}^{r-1}({\cal K}_{l,l+1}+2{\cal I}_{l,l+1}-2{\cal
P}_{l,l+1})~:V^{\tens r}\riya V^{\tens r}}
where the two ends do not interact is called an open spin chain
hamiltonian\foot{See \refs{\stefan,\cheww,\dewman} for recent
works on the open spin chain description of anomalous dimensions in
certain gauge theories.}.  An integrable open spin chain is described
by an integrable closed spin chain interaction in the bulk together with
boundary conditions at the two ends denoted by $K^\pm$.  In order for
the resulting open chain to be integrable, $K^\pm$ must
satisfy the boundary Yang-Baxter equations \sklyan
\eqn\Bybe{\eqalign{&R_{12}(u-v)K^-_1(u)R_{12}(u+v)K^-_2(v) =
K^-_2(v)R_{12}(u+v)K^-_1(u)R_{12}(u-v)\cr 
&R_{12}(-u+v)K^{+t_1}_1(u)R_{12}(-u-\!v-\!2\eta)K^{+t_2}_2(v) \cr &\hskip40mm =
K^{+t_2}_2(v)R_{12}(-u-\!v-\!2\eta)K^{+t_1}_1(u)R_{12}(-u+v)}}
where for $SO(6)$ spin chain, $\eta=-2$ and the $R$-matrix is given by
\eqn\rmatrix{R_{12}(u)=\frac{1}{2}\left[u(u-2)\I_{12}-(u-2)\PP_{12}
+u{\cal K}_{12}\right]~.}
$t_i~(i=1,2)$ in \Bybe\ denotes the transpose on the $i^{\rm th}$
vector space.  The hamiltonian in \Hs\ corresponds to the ``free'' boundary
conditions $K^\pm=1$ which trivially satisfies \Bybe.  

Having identified open spin chain hamiltonians in the definitions of
$b_{ij}$, one may wonder how open spin chain states can arise from
closed spin chain states.  Simply, we can split a closed spin chain
into two, thereby breaking the periodicity, to give
two associated open spin chains.  In this respect, operators ${\cal
O}_{1,2,3}$ can be regarded as matrix operators acting on the open
spin chain Hilbert space as follows.  Recalling that the lengths of
operators and $r,s,t$ in \rst\ are related by
\eqn\Lirst{L_1=r+s,\quad L_2=s+t, \quad L_3=t+r,}
we define the open spin chain matrix operators $\Psi_{1,2,3}$
corresponding to operators ${\cal O}_{1,2,3}$ by
\eqn\psiLR{\eqalign{
\Psi_{1}&=\tilde{\psi}_{i_{31}\ldots i_{3r}i_{2s}\ldots i_{21}}
|i_{31}\ldots i_{3r}\ket\bra i_{21}\ldots i_{2s}|
\quad~:V^{\tens s}\riya V^{\tens r} \cr
\Psi_{2}&=\tilde{\psi}_{j_{11}\ldots j_{1s}j_{3t}\ldots j_{31}}
|j_{11}\ldots j_{1s}\ket\bra j_{31}\ldots j_{3t}| 
~~~:V^{\tens t}\riya V^{\tens s}\cr
\Psi_{3}&=\tilde{\psi}_{k_{21}\ldots k_{2t}k_{1r}\ldots k_{11}}
|k_{21}\ldots k_{2t}\ket\bra k_{11}\ldots k_{1r}| 
\,\,:V^{\tens r}\riya V^{\tens t}.
}}
The ket $|i_{31}\ldots i_{3r}\ket$ and the bra $\bra i_{21}\ldots i_{2s}|$
represent the open spin chain states associated with the closed spin chain 
${\cal O}_1$.  In terms of these spin chain operators, $c^1_{123}$,
as given by \ctwoexp\ with \fbdef, is written simply as
\eqn\conecompact{\la c^1_{123}=-\hf\sum_{i=1}^3\ga_i+{\la\o16\pi^2}
{\Tr_r(H_r\Psi_1\Psi_2\Psi_3+\Psi_1H_s\Psi_2\Psi_3+\Psi_1\Psi_2H_t\Psi_3)
\o \Tr_r \Psi_1\Psi_2\Psi_3}.}
Above, $\Tr_r$ is the trace over the vector space $V^{\tens r}$
and $H_r$ is the hamiltonian acting on a open spin chain of length $r$
as in \Hs.

Equation \conecompact\ can be further simplified by introducing
the following matrices $M_{r,s,t}$
\eqn\Mrst{
M_r={\Psi_1\Psi_2\Psi_3\o\Tr_r\Psi_1\Psi_2\Psi_3},\quad
M_s={\Psi_2\Psi_3\Psi_1\o\Tr_s\Psi_2\Psi_3\Psi_1},\quad
M_t={\Psi_3\Psi_1\Psi_2\o\Tr_t\Psi_3\Psi_1\Psi_2}.
}
In terms of these matrices, \conecompact\ is rewritten as
\eqn\conerho{\eqalign{
\la c^1_{123}&=-\hf\sum_{i=1}^3\ga_i+{\la\o16\pi^2}
\sum_{k=r,s,t}\Tr_k(H_kM_k)\cr
&=-\hf\sum_{i=1}^3\ga_i+{\la\o16\pi^2}
\sum_{k=r,s,t}\Tr_k(H_k\rho_k)}}
where $\rho_k$ is the symmetric part of $M_k$
\eqn\rhokMk{\rho_k=\hf(M_k+M_k^T),}
and $M_k^T$ denotes the transpose of $M_k$.
Note that from \Mrst\ and \rhokMk,
$\rho_k$ satisfies
\eqn\Trrhok{\Tr_k(\rho_k)=1.}
From this property and the fact that
$\rho_k$ is a real symmetric matrix (see footnote 7), we interpret $\rho_k$
as a ``density matrix'' and the second term of $c^1_{123}$ in
\conerho\ as the ensemble average of open chain hamiltonian $H_k$
with weight $\rho_k$.  The appearance of density matrices seems natural
from our definition of open chain states.  By the construction of open
spin chain states as the two segments of one closed spin chain,
the resulting open spin states are correlated, and in some sense,
they can be thought of as ``mixed states.''

The integrability of open spin chain hamiltonian
$H_k$ could be useful to compute $c^1_{123}$ given in \conerho.
By utilizing the powerful techniques of integrable system, such as 
the algebraic Bethe ansatz, we can in principle diagonalize $H_k$ as
\eqn\diagH{
H_k=\sum_{\al=1}^{6^k}{\cal E}_{\al}^{(k)}|v_{\al}^{(k)}\ket\bra v_{\al}^{(k)}|.}
Here, we orthonormalized the eigenvectors: 
$\bra v_{\al}^{(k)}|v_{\bt}^{(k)}\ket=\cob_{\al\bt}$.
Once the eigenvalues and eigenvectors of $H_k$ are known,
$\Tr_k(H_k\rho_k)$ appearing in \conerho\ can be written as 
the average of eigenvalues with respect to
the ``probability'' $p_{\al}^{(k)}=\bra
v_{\al}^{(k)}|\rho_k|v_{\al}^{(k)}\ket$
\eqn\trHkpk{
\Tr_k(H_k\rho_k)=\sum_{\al=1}^{6^k}{\cal E}_{\al}^{(k)}p_{\al}^{(k)},}
where $\sum_{\al}p_{\al}^{(k)}=1$. 

In our spin chain interpretation of $c^1_{123}$, we have utilized
the connection between $b_{ij}$ and open spin chain hamiltonians.
In contrast, for $f^i_{jk}$, its $SO(6)$ index structure as given in \fbdef\ does not directly lend itself to an interpretation in terms of
conventional spin chain language.  We can however use the
relations between two- and three-point constants in \aeq\ to obtain a
heuristic understanding of $f^i_{jk}$.   For instance, \aeq\ suggests
that we can write $f^1_{23}$ as the energy difference between a closed
spin chain and its associated open spin chains
\eqn\fasEdiff{
f^1_{23}=-{\bra\Psi_1|H|\Psi_1\ket\o\bra\Psi_1|\Psi_1\ket}
+\Tr_r(H_r\rho_r)+\Tr_s(H_s\rho_s).}
In other words, $f^1_{23}$ represents roughly 
the ``energy cost'' in splitting a closed chain of length $L_1=r+s$  
into the associated open chains of lengths $r$ and $s$.    
Since $c^1_{123}=\frac{1}{32\pi^2}\left(\f+\ff+\fff\right)$, 
$c^1_{123}$ as given in \conerho\ can also be interpreted
as the total energy cost in splitting all three closed
chains.  As for the extremal case where
$\la c^1_{123}=\hf(\ga_2+\ga_3-\ga_1)$, $c^1_{123}$ has a clear
interpretation  as the energy cost in the
splitting of a single closed chain into two separate closed chains.

\fig{Schematic diagram depicting one of the terms in $c^1_{123}$. 
$H_t$ is the open spin chain hamiltonian. 
(In this figure $r=2$, $s=t=3$.) The arrows indicate
the orientation we choose in order to write ${\cal O}_{1,2,3}$ as 
matrix operators $\Psi_{1,2,3}$.  $c^1_{123}$ is clearly independent 
of the choice of orientation.}
{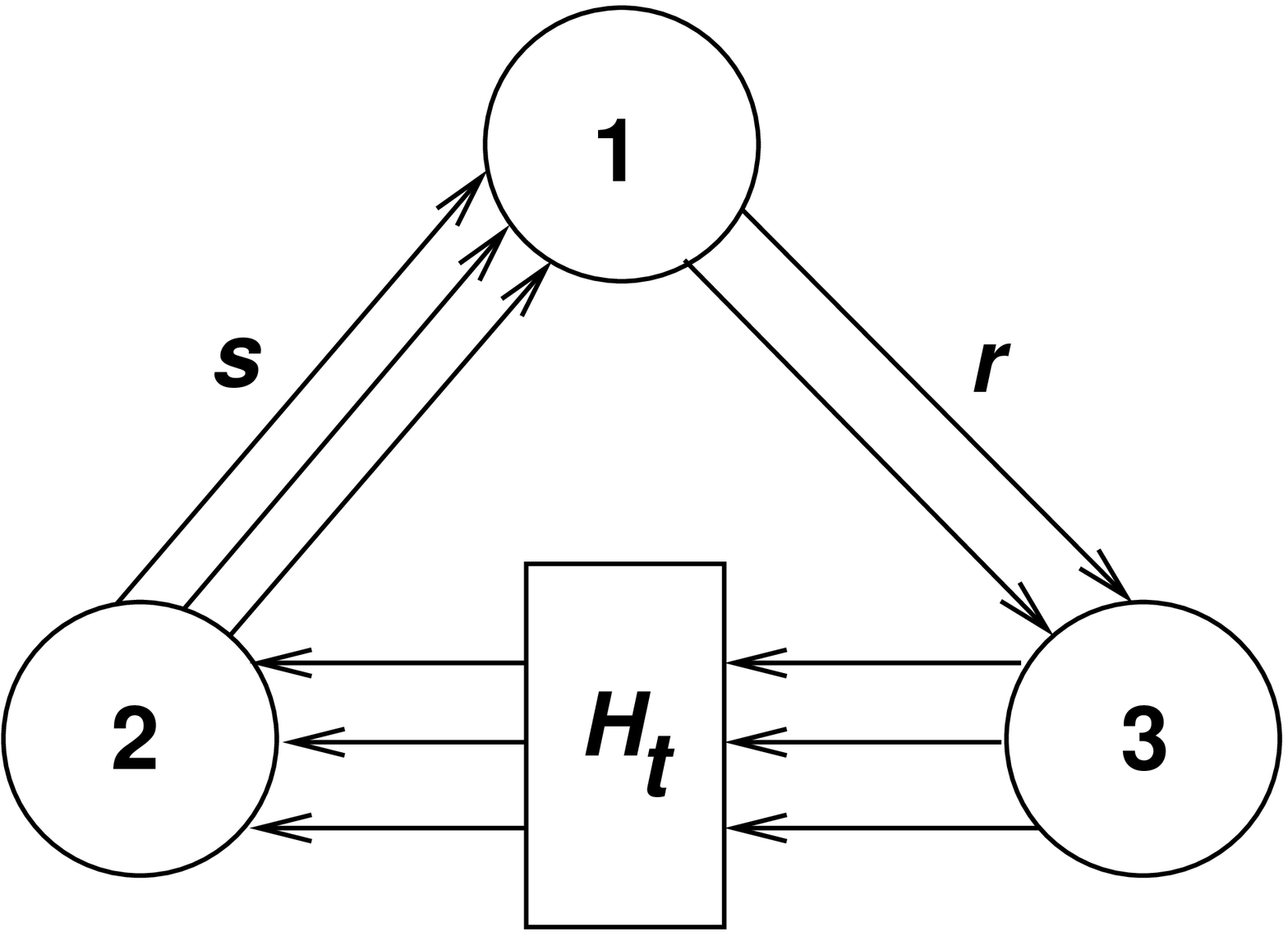}{40mm}
In passing, we note that the two terms in $\rho_k$ \rhokMk\ have an
interesting interpretation in the diagrammatic representation of
$c^1_{123}$, as shown in Fig.\ 7.  In writing operators ${\cal O}_{1,2,3}$
as matrix operators $\Psi_{1,2,3}$, we implicitly assumed that the diagram
describing the index contraction has a definite orientation, which corresponds
to the first term $M_k$ in \rhokMk.
On the other hand, the diagram for $M_k^T$ carries the opposite orientation
from that of $M_k$.  Equations \conerho\ and \rhokMk\ imply that the two
orientations contribute to $c^1_{123}$ with equal weight.  This is
consistent with the fact that the three-point function coefficient
$c_{123}$ is symmetric with respect to the indices $1,2,3$.

\subsec{Some Examples}

Let us illustrate our general formula \conecompact\ with some 
examples of operators with small scaling dimensions.
We list the eigen-operators consisting up to three scalar fields,
\eqn\lowop{\eqalign{
K&=\frac{1}{\lam}\,\tr\phi^m\phi^m\cr
{\cal O}_{(ij)}&=\frac{1}{\lam}(\tr\phi^i\phi^j-\cob^{ij}{1\o6}\tr\phi^m\phi^m)\cr
K_i&=\frac{1}{\lam^{3/2}}\,\tr\phi^m\phi^m\phi^i \cr
{\cal O}_{(123)}&=\frac{1}{\lam^{3/2}}(\tr\phi^1\phi^2\phi^3+\tr\phi^3\phi^2\phi^1)\cr
{\cal O}_{[123]}&=\frac{1}{\lam^{3/2}}(\tr\phi^1\phi^2\phi^3-\tr\phi^3\phi^2\phi^1).
}}
${\cal O}_{(123)}$ belongs to the symmetric traceless representation
of $SO(6)$ and ${\cal O}_{[123]}$ belongs to the anti-symmetric representation.
Their anomalous dimensions are given by
\eqn\amonvarious{
\ga_{{\cal O}_{(ij)}}=\ga_{{\cal O}_{(123)}}=0,\quad
\ga_K=\ga_{{\cal O}_{[123]}}={3\o4\pi^2}\la,\quad
\ga_{K_i}={1\o2\pi^2}\la.}
And note that ${\cal O}_{(ij)}$ and ${\cal O}_{(123)}$ are $1/2$ BPS
operators.

As a first example, let us reconsider $c^1_{KKK}$ (studied in section 3)
from the general formula \conecompact.  The matrix associated with the
Konishi operator is $\Psi_K=2\sum_{m=1}^6 |m\ket\bra m|$.
Since the open spin hamiltonian $H_1$ acting on length 1 chain is zero,
there is no contribution from the second term in \conecompact.
Therefore, $\la c^1_{KKK}$ is given by $-{3\o2}\ga_K=-{9\o8\pi^2}\la$,
in agreement with our computation in section 3.  Next, let us consider
the slightly more complicated example $c^1_{{\cal O}_{(12)}K_1K_2}$.
The matrices associated with operators ${\cal O}_{(12)}, K_1$, and $K_2$
are given by
\eqn\PsiOij{\eqalign{
\Psi_{{\cal O}_{12}}&=|1\ket\bra 2|+|2\ket\bra 1|\cr
\Psi_{K_1}&=\sum_{m=1}^6|1\ket\bra m,m|+|m\ket\bra m,1|+|m\ket\bra 1,m|\cr
\Psi_{K_2}&=\sum_{m=1}^6|m,m\ket\bra 2|+|m,2\ket\bra m|+|2,m\ket\bra m|.
}}
Again, the open spin chain hamiltonian acting on the length 1 part is
zero.  The only nontrivial contribution comes from the length 2 part.
The density matrix $\rho_2$ associated with \PsiOij\ is given by
\eqn\exrhotwo{\eqalign{
\rho_2=&
{1\o24}\Big(|K\ket+2|1,1\ket\Big)
\Big(\bra K|+2\bra 2,2|\Big)
+{1\o24}\Big(|K\ket+2|2,2\ket\Big)
\Big(\bra K|+2\bra 1,1|\Big) \cr
+&{1\o12}\Big(|1,2\ket+|2,1\ket\Big)\Big(\bra 1,2|+\bra 2,1|\Big)
}}
where $|K\ket=\sum_{m=1}^6|m,m\ket$.
Plugging $\ga_{{\cal O}_{(12)}}=0$, $\ga_{K_i}={\la\o2\pi^2}$ 
and $\Tr_2(H_2\rho_2)={16\o3}$
into \conerho, $c^1_{{\cal O}_{(12)}K_1K_2}$
is found to be
\eqn\coneOKK{
\la c^1_{{\cal O}_{(12)}K_1K_2}=-\hf(\ga_{K_1}+\ga_{K_2})
+{\la\o 16\pi^2}\Tr_2(H_2\rho_2)
=-{1\o6\pi^2}\la~.
}
One can easily check that the conformal constraints \garel\
are satisfied for this example.

In a similar manner, we can compute various 
$c^1_{123}$ for operators in \lowop.
Below we list the non-vanishing combinations of $c^1_{123}$
\eqn\Cs{\eqalign{
&c^1_{K{\cal O}_{(ij)}{\cal O}_{(ij)}}=-{3\o8\pi^2}\cr
&c^1_{KK_iK_i}=-{17\o24\pi^2}\cr
&c^1_{{\cal O}_{(ij)}K_iK_j}=-{1\o6\pi^2}\cr
}\qquad
\eqalign{
&c^1_{K{\cal O}_{(123)}{\cal O}_{(123)}}=-{3\o8\pi^2}\cr
&c^1_{K{\cal O}_{[123]}{\cal O}_{[123]}}=-{7\o8\pi^2}\cr
&c^1_{{\cal O}_{(12)}{\cal O}_{[134]}{\cal O}_{[234]}}=-{1\o2\pi^2}~.
}}
We can also check that 
the following combinations vanish as expected from the non-renormalization
theorem
\eqn\nonRcs{
c^1_{{\cal O}_{(12)}{\cal O}_{(23)}{\cal O}_{(31)}}
=c^1_{{\cal O}_{(12)}{\cal O}_{(134)}{\cal O}_{(234)}}=0~.
}

Another interesting set of operators is the BMN operators \BMN.
The exact two impurity BMN operators which diagonalize
the planar one-loop dilatation operator were constructed by Beisert \BeisertTN
\eqn\BMNbei{
{\cal B}_{(34)}^{J,n}=\frac{1}{\lam^{J/2+1}}\sum_{k=0}^J\cos{\pi n(2k+1)\o J+1}\tr\phi^3 Z^k\phi^4 
Z^{J-k}~,
}
where  $Z=\phi^1+i\phi^2$. 
The one-loop anomalous dimension of this operator is 
$\ga_{{\cal B}_{(34)}^{J,n}}={\la\o\pi^2}\sin^2\left[{\pi n\o J+1}\right]$. 
Vanishing of the anomalous dimension for $n=0$ reflects the fact that 
${\cal B}_{(34)}^{J,n=0}$ is a 1/2 BPS operator.
Using our formula \conecompact, we compute some examples of
$c^1_{123}$ involving the BMN operators:
\eqn\BMNga{\eqalign{
c^1_{{\cal O}_{(12)}{\cal B}_{(34)}^{J,n}\overline{{\cal B}}_{(34)}^{J,n}}
&=-{1\o2\pi^2}\sin^2{\pi n\o J+1}\cr
c^1_{K{\cal B}_{(34)}^{J,n}\overline{{\cal B}}_{(34)}^{J,n}}&=
-{3\o8\pi^2}-{2\o\pi^2}{1\o J+2}\sin^2{\pi n\o J+1}
\lf(1-{1\o J+1}\cos^2{\pi n\o J+1}\ri).
}}
It is interesting to note that $c^1_{({\rm Konishi})({\rm BPS})({\rm BPS})}$
is independent of the choice of BPS operators,
\eqn\cKSS{
c^1_{K{\cal O}_{(12)}{\cal O}_{(12)}}=
c^1_{K{\cal O}_{(123)}{\cal O}_{(123)}}
=c^1_{K{\cal B}_{(34)}^{J,0}\overline{{\cal B}}_{(34)}^{J,0}}
=-{3\o8\pi^2}.}

\newsec{Relation to Closed String Field Theory}

According to the AdS/CFT correspondence,
a $n$-point correlation function of single trace operators
in ${\cal N}=4$ SYM theory corresponds to
an on-shell $n$-particle amplitude in the bulk
type IIB string theory on $AdS_5\times S^5$.
The perturbative SYM theory with small 't Hooft coupling $\la$
is dual to the bulk string theory in highly curved background with
curvature radius $R_{AdS}\sim\la^{\qu}\rt{\al'}$.
In this regime of coupling, it is very difficult to
compare the SYM results with those of string theory,
because the worldsheet theory is a strongly coupled nonlinear sigma
model with RR flux.  One exception is the sector of SYM carrying large
R-charge which corresponds to the string theory on a pp-wave
background geometry \BMN.  The worldsheet theory in pp-wave background
becomes free in the light-cone gauge,
and it was shown that the spectrum on the bulk side
is reproduced by the so-called BMN operators in the SYM side \BMN.
Moreover, the string interactions described by
the light-cone string field theory \SpradlinAR\ 
agree with the three-point functions on the SYM side.
In the light cone gauge, the light-cone momentum is conserved
$p_1^+=p_2^++p_3^+$.
On the SYM side, this corresponds to the conservation of
the length of spin chains $L_1=L_2+L_3$.
Namely, the three-string interactions in the light-cone gauge are
dual to the extremal three-point functions.

For general correlators in ${\cal N}=4$ SYM theory outside of the BMN sector, 
there is currently no string theory result to compare.  
Below, we speculate on some features of string interactions in
$AdS_5\times S^5$ based on our computation of three-point functions in
the SYM side.  It is natural to expect that the non-extremal
correlators in the SYM side correspond to the covariant three string interactions in $AdS_5\times S^5$.
In fact, the pattern of index contraction in free SYM theory
in Fig.\ 4 fits nicely with the Witten type vertex in closed string
field theory as shown in Fig.\ 8 \refs{\WittenSFT,\GiddingsBP}.
This vertex is a natural generalization of the mid-point interaction
vertex in open string field theory.  There are two special points
$P_{1,2}$ in this three-string vertex which are the counterpart 
of the mid-point in open string field theory. 
The points $P_{1,2}$ divide each closed string into two segments
with lengths $r,s,t$ given by \rst\ and the closed strings are glued
along these segments.  One can easily see the correspondence between
the ``triangular'' type diagram of free planar three-point
contractions in Fig.\ 4 and the Witten type three-string overlap
vertex in Fig.\ 8.  The point $P_1$ and $P_2$ corresponds respectively
to the area inside and outside the ``triangle'' in Fig.\ 4.  

\fig{Witten type three-string vertex in closed string field theory. 
The two interaction points $P_{1,2}$ are the analogues of the mid-point
in open string field theory.}
{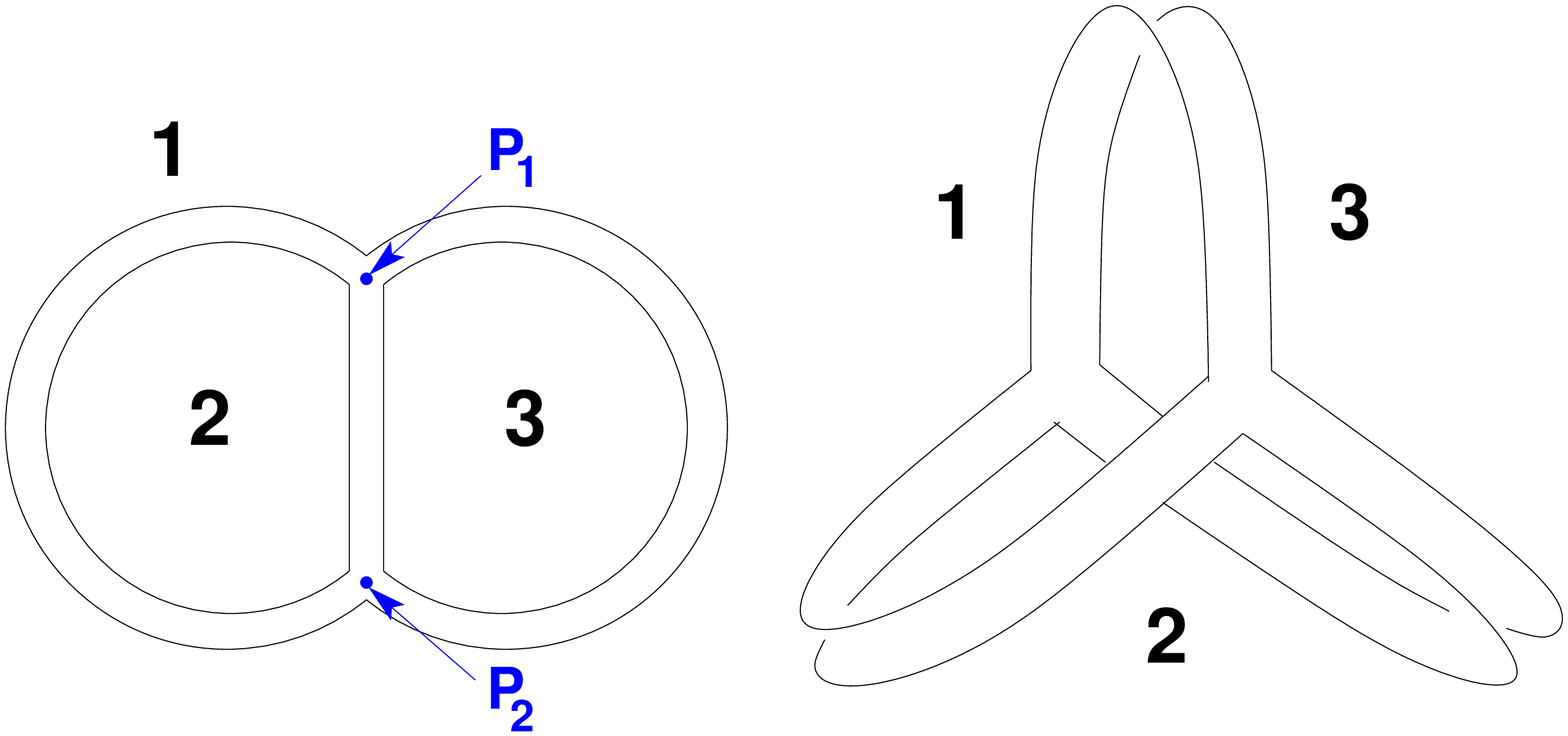}{80mm}
\fig{Zoom-up of one of the two interaction points $P_{1,2}$. 
(a) Three-point interactions $f^i_{jk}$ are localized at the
interaction points.  (b) Two-point interactions $b_{ij}$ are
along the overlapping segments.}
{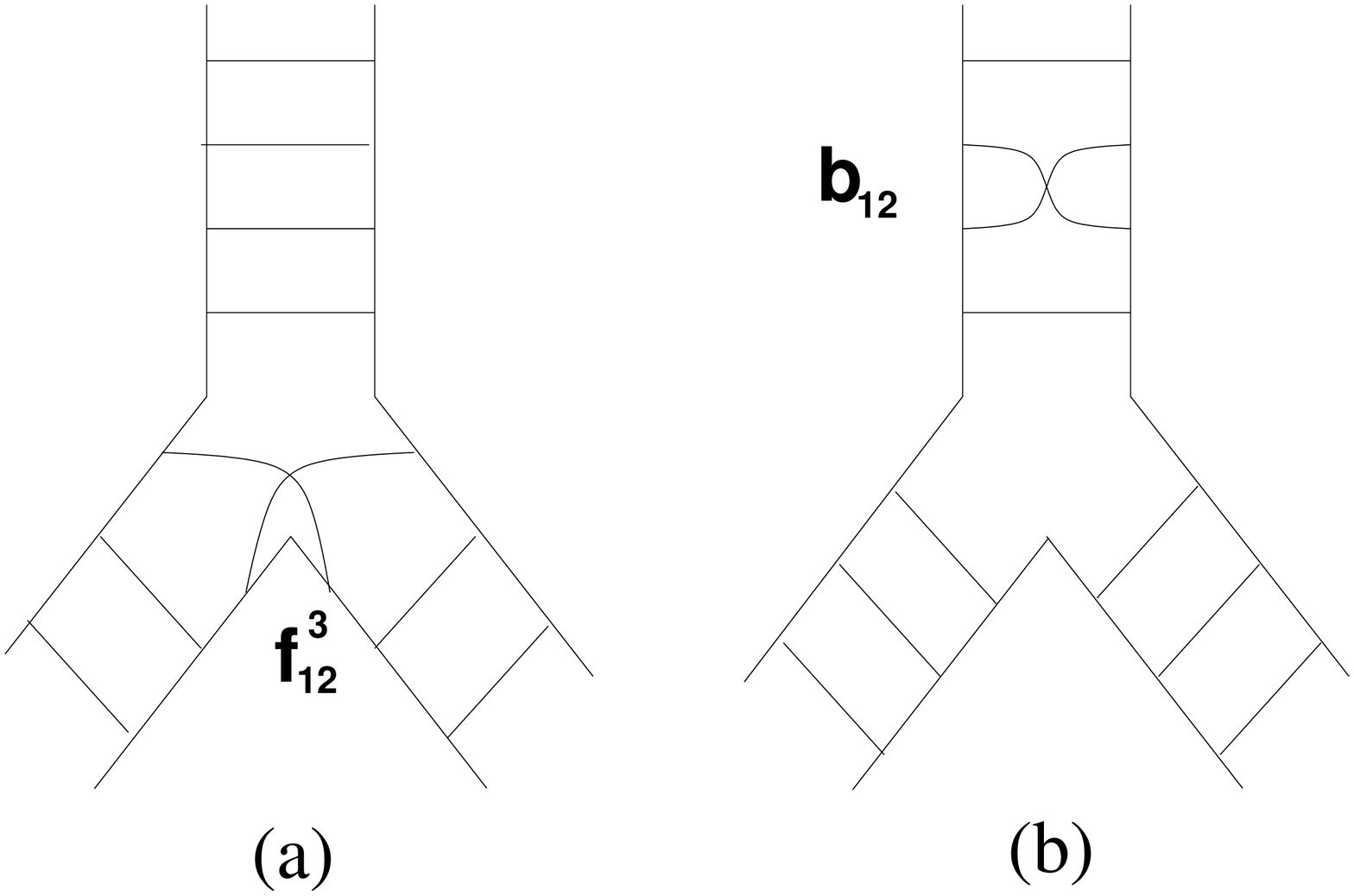}{70mm}

We found that the structure constant $c_{123}$ gets corrected from the
free theory value by the amount 
$\la\, c^1_{123}={\la\o32\pi^2}\lf(\f+\ff+\fff\ri)$
when we turned on the 't Hooft coupling $\la$.
The two terms in the definition of $f^i_{jk}$ in \fbdef\ 
corresponds to the two interaction points $P_{1,2}$
and both have the form of the hamiltonian, 
${\cal K}+2{\cal I}-2{\cal P}$, acting on one link at the interaction
point, as shown in Fig.\ 9a.  This reminds us of the mid-point insertion of
operators in open string field theory.  In Witten's open
string vertex, the only natural place to insert an operator
without breaking (naive) associativity is the mid-point. 
We expect that $P_{1,2}$ are the natural points to insert an operator 
also for the closed string case,
although the product of closed string fields is non-associative.
In the bulk string theory side, turning on the 't Hooft coupling $\la$ 
corresponds to changing the radius of $AdS_5\times S^5$ by
adding some operator $\la\int d^2z{\cal V}$ to the worldsheet action $S$.
The form of $c_{123}^1$ suggests that the closed three-string vertex 
$|V_3^{(\la=0)}\ket$ at $\la=0$ is modified by 
$\la\Big(\til{{\cal V}}(P_1)+\til{{\cal V}}(P_2)\Big)|V_3^{(\la=0)}\ket$,
where $\til{{\cal V}}$ is some vertex operator on the worldsheet
which can be thought of as the continuum version of $f^i_{jk}$.
It would be interesting to know the precise relation between
${\cal V}$ and $\til{{\cal V}}$.  Alternatively, the SYM theory provides
us with a complementary description of this deformation.  The
alternative expression \ctwoexp\ implies that
instead of inserting an operator at $P_{1,2}$, we can insert
open spin chain like interactions in the bulk of closed string
worldsheet and treat $P_{1,2}$ as defects.  
This is shown in Fig. 9b.
To our knowledge, this type of deformation of three-string vertex has
not been studied.  We suspect that supersymmetry and conformal symmetry
will again play an important role for understanding this complementarity
from the string theory point of view.

\newsec{Discussions}
In this paper, we have calculated the one-loop correction to the
three-point function coefficient in ${\cal N}=4$ SYM theory.  In
general, this is not identical to the $O(\lam)$ correction.  At the
one-loop level, operators mixes with coefficients independent of
$\lam$. In particular, the diagonalization problem of the spin chain
hamiltonian \ham\ does not involve $\la$.
However, at the two-loop level, this is no longer the case. 
The anomalous dimension matrix is now $O(\lam^2)$, and  
correspondingly, the eigen-operators may become $\la$ dependent at this
order: $\O_\al=\O_\al^{(0)} +\la\O_\al^{(1)}+O(\la^2)$.
This $\la$ dependent term in the eigen-operator affects 
the order $\lam$ correction to the three-point coefficient
through the free contraction such as 
$\la\bra\O_\al^{(1)}\O_\bt^{(0)}\O_\rho^{(0)}\ket_{\rm free}$.  
We emphasize that this correction is a two-loop effect which we do
not consider in this paper.  With regards to planar integrability, this
suggests that $O(\lam)$ integrability requires two-loop integrability.

However, for one-loop planar integrability, the one-loop correction
is the natural one to consider as it directly relates to integrable
spin chains.  The eigenstates of the spin chain hamiltonian (i.e. the
first order anomalous dimension matrix) correspond to the eigen-operators we
considered in this paper.  Interestingly, the calculation of the one-loop 
correction to the three-point coefficient suggests a new mechanism for
closed spin chains - that is, a chain can split into two chains, or
conversely, that two chains can ``glue'' together to form a third
chain.  Although such processes seem natural from the field theory
point of view, to our knowledge, such notions have not been studied within
the context of spin chains.  The extremal case would correspond to the gluing
and splitting process that conserve the total number of spin sites.  And
the energy cost of such processes would be provided by the one-loop
three-point function correction.  For the non-extremal case, the total
number of spin sites would not be conserved.

As for planar \Nfour SYM theory, we have shown that one-loop
correction for both anomalous dimension and three-point coefficient
can be calculated using spin chains.  We have however not exploited
integrable techniques for open spin chain that may possibly be useful for
calculating non-extremal three-point coefficients of large scaling
dimension operators.  Because of the presence of the density matrices,
it is not clear to us what is the most efficient manner to utilize
integrability.  

The usefulness of the integrable $SO(6)$ spin chain for planar \Nfour SYM
derives from the fact that the values of interest thus far are all
solely dependent on the $SO(6)$ indices of the scalar fields.  It
would be interesting to see whether the chains of indices (or the spin chain
data) by themselves are sufficient to directly
determine the planar four- or higher-point functions whose spatial
dependence is not completely constrained by conformal symmetry.  From
the point of view of AdS/CFT correspondence, $n$-point functions in
the SYM side correspond to $n$-string interactions in $AdS_5\times
S^5$.  In the covariant closed string field theory of non-polynomial
type \refs{\Saadi,\Kugo}, it is known that $n$-string interaction
vertices with arbitrarily high $n$ are required in order to cover the
whole moduli space of closed Riemann surface with punctures.  It would
be interesting to relate this with the structure of $n$-point
functions in SYM theory.

There are additionally several directions to extend our work.
One natural extension is to calculate
three-point functions of more general operators outside the $SO(6)$ subsector.
The original result of Minahan and Zarembo for the planar one-loop anomalous
dimension in the $SO(6)$ subsector has been extended to the general
$PSU(2,2|4)$ sector in \refs{\beiks,\bei,\beis}. 
It would be interesting to repeat our analysis of the three-point functions
for the $PSU(2,2|4)$ sector and see if a similar role is played by
an analogous $PSU(2,2|4)$ open spin chain.  It would also be
interesting to study the higher loop corrections to the three-point
function coefficients and their relations to integrable spin chains.

\bigskip\bigskip
\centerline{\bf Acknowledgements}
We are grateful to K.~Becker, J.~Harvey, and Y.~S.~Wu for their
interest in this work and many helpful discussions.  We also would
like to thank J.~Dai and M. Dine for useful conversations.  K.O. is
supported in part by DOE grant DE-FG02-90ER40560.  L.S.T. is supported
in part by NSF grants PHY-0204608 and PHY-0244722.

\appendix{A}{Computation of $B$ and $F$}

The function $B(\x,\xx)$ as defined in \twofeyn\ is given by
\eqn\Bexp{\eqalign{B(\x,\xx)& = 2 \lam \int d^4y\,
\frac{[8\pi^2|x_{12}|^2]^2}{[8\pi^2|y-\x|^2]^2\,[8\pi^2|y-\xx|^2]^2}
\cr & = \frac{2 \lam}{(8\pi^2)^2} |x_{12}|^4 \int  d^4y\, \frac{1}{|y-\x|^4
|y-\xx|^4}} } 
We regularize the propagator by differential regularization
\frejmv\ where a small distance cutoff is inserted into the
propagator.  Explicitly,  
\eqn\lapinfdif{
\lap(y-x)={1\o(y-x)^2+\ep^2}~.
}
The integral in \Bexp\ is easily evaluated as
\eqn\Bint{
\int d^4y\;\lap(y)^2\lap(y-x)^2
=\pi^2\int_0^1 dt\,{ t(1-t)\o[t(1-t)x^2+\ep^2]^2} 
={2\pi^2\o x^4}\Big[\log{x^2\o\ep^2}-1\Big]~.
}
Letting $\Lam^2=1/\ep^2$, this gives
\eqn\Bdiff{B(\x,\xx) = \frac{\lam}{16\pi^2}\left[\ln|{x_{12}\Lam}|^2-1\right]~.}

$F(\xxx;\x,\xx)$ is defined in \threefeyn\ and takes the form
\eqn\Fexp{\eqalign{F(\xxx;\x,\xx)&= 2\lam \int d^4y\,
\frac{[8\pi^2|x_{13}|^2][8\pi^2|x_{23}|^2]}{[8\pi^2|y-\xxx|^2]^2\,[8\pi^2|y-\x|^2][8\pi^2|y-\xx|^2]}\cr
&=\frac{2\lam}{(8\pi^2)^2} |x_{13}|^2|x_{23}|^2 \int d^4y\,
\frac{1}{|y-\xxx|^4|y-\x|^2|y-\xx|^2} ~.} }
Using differential regularization, we need to calculate the integral
$I_3$ defined by
\eqn\intCdef{
I_3=\int d^4y\; \lap(y)^2\lap(y-x_1)\lap(y-x_2)~.
}
To compute $I_3$, it is useful to use the inversion
\eqn\inversion{x^\mu\riya\h{x}^\mu={x^\mu\o x^2}~.} 
Obviously, this map is involutive $\h{\h{x^\mu}}=x^\mu$,
and the norm is transformed as
\eqn\norminv{
(\h{x})^2={1\o x^2}, \quad
(\h{x}_1-\h{x}_2)^2={(x_1-x_2)^2\o x_1^2x_2^2}~.
}
The measure and the propagator transform as
\eqn\dzinv{
d^4\h{z}={d^4z\o(z^2)^4}, \quad
\lap(\h{z}-x)=\lap(\h{z}-\h{\h{x}})
={1\o x^2}{z^2\o (z-\h{x})^2+\ep^2z^2\h{x}^2}~.
}
By changing the integration variable 
$y\riya \h{z}$, the integral $I_3$ becomes
\eqn\Cint{\eqalign{
I_3&=\int d^4\h{z}\; \lap(\h{z})^2\lap(\h{z}-x_1)\lap(\h{z}-x_2) \cr
&={1\o x_1^2x_2^2}\int d^4z\; {1\o(1+\ep^2 z^2)^2}
{1\o (z-\h{x}_1)^2+\ep^2z^2\h{x}_1^2}
{1\o (z-\h{x}_2)^2+\ep^2z^2\h{x}_2^2}~.
}}
Observing that the logarithmic divergence is coming from
$z=\infty$, up to ${\cal O}(\ep)$ we can replace $I_3$ by
\eqn\Csimp{\eqalign{
I_3&={1\o x_1^2x_2^2}\int d^4z\; {1\o(1+\ep^2 z^2)^2}
{1\o z^2}
{1\o (z-(\h{x}_1-\h{x}_2))^2} \cr
&={1\o x_1^2x_2^2}\int d^4z\; {1\o(1+ z^2)^2}
{1\o z^2}
{1\o (z-\ep(\h{x}_1-\h{x}_2))^2}~.
}}
Using Feynman parameter, $I_3$ is written as
\eqn\Cparam{\eqalign{
I_3&={\pi^2\o x_1^2x_2^2}
\int_0^1dt\int_0^{1-t}ds\;s[s+t(1-t)\ep^2(\h{x}_1-\h{x}_2)^2]^{-2}\cr
&={\pi^2\o x_1^2x_2^2}\int_0^1dt\lf[\log\lf({1+t\ep^2(\h{x}_1-\h{x}_2)^2
\o t\ep^2(\h{x}_1-\h{x}_2)^2}\ri)
-{1\o 1+t\ep^2(\h{x}_1-\h{x}_2)^2}\ri] \cr
&=-{\pi^2\o x_1^2x_2^2}\log \ep^2(\h{x}_1-\h{x}_2)^2+{\cal O}(\ep) \cr
&={\pi^2\o x_1^2x_2^2}\log{x_1^2x_2^2\o \ep^2(x_1-x_2)^2}~.
}}
Applying this result into \Fexp\ and again taking $\Lam^2=1/\ep^2$, we obtain
\eqn\Fdiff{
F(\xxx;\x,\xx)=\frac{\lam}{32\pi^2} 
\ln\lf|{x_{13}x_{23}\La\o x_{12}}\ri|^2~.
}

Introducing the renormalization scale $\mu$, we can renormalize $B$
and $F$ for example by taking $\mu=\Lam$,
\eqn\BFdef{
B(x_1,x_2)=b_0+{\la\o16\pi^2}\ln|x_{12}\mu|^2,\quad
F(x_3;x_1,x_2)=f_0+{\la\o32\pi^2}\ln\lf|{x_{13}x_{23}\mu\o x_{12}}\ri|^2~.
}
where $b_0=-\frac{\lam}{16\pi^2}$ and $f_0=0$ for differential
regularization but it is useful here to leave it arbitrary.  Under the
rescaling $\mu\riya\mu e^a$, $b_0$ and $f_0$ are shifted as follows,
\eqn\bfshift{
b_0'=b_0+{\la a\o8\pi^2}~,\quad f_0'=f_0+{\la a\o16\pi^2}~.
}
Clearly, the combination $2f_0'-b_0'=2f_0-b_0$ is independent of the
renormalization scale of $\mu$.  Thus, from the explicit calculation,
\eqn\invfbcomb{2f_0-b_0=\frac{\lam}{16\pi^2} }
is scheme independent.

\appendix{B}{Index Contraction as a Matrix Integral}
The free contraction coefficient $\tilde{C}^0$ appeared in the text
can be written as a large $N$ planar matrix integral
\eqn\matrixC{
\Big\bra\Big\bra\Tr( X^{i_1}\cdots X^{i_{L_1}})
\Tr (Y^{j_1}\cdots Y^{j_{L_2}})
\Tr (Z^{k_1}\cdots Z^{k_{L_3}})\Big\ket\Big\ket_{{\rm planar}}
}
where the matrix model expectation value $\bra\bra{\cal O}\ket\ket$
is defined by
\eqn\XYZaction{\eqalign{
\bra\bra{\cal O}\ket\ket&=\int\prod_{m=1}^6dX^mdY^mdZ^m\;
{\cal O}e^{-S_{\rm 3pt}},\cr
S_{\rm 3pt}&=\frac{1}{4}\sum_{m=1}^6
\Tr\Big[(X^m)^2+(Y^m)^2+(Z^m)^2-2X^mY^m-2Y^mZ^m
-2Z^mX^m\Big]~.
}}
$X^m,Y^m$ and $Z^m$ are $N\times N$ hermitian matrices.
The propagator of these matrices derived from the action \XYZaction\
is
\eqn\propXYZ{
\bra\bra (X^m)^a_b(Y^n)^c_d\ket\ket=
\bra\bra (Y^m)^a_b(Z^n)^c_d\ket\ket=
\bra\bra (Z^m)^a_b(X^n)^c_d\ket\ket=\cob^{m,n}\cob^a_d\cob^c_b~,
}
where $a,b,c,d$ are the color indices.
These propagators connect the matrices in \matrixC\
without the self-contraction inside the trace.
In the planar limit, the resulting contraction is exactly the same
as that appears in the computation in the free gauge theory.

The one-loop corrected three-point function coefficient would be obtained
by adding interactions to the matrix model.  Note that the one loop corrected 
dilatation operator obtained in \beiks\ can be written as a matrix model
\eqn\anommatrix{\eqalign{
&\Ga^{i_1\cdots i_L}_{j_1\cdots j_L}
=\int\prod_{m=1}^6dX^mdY^m\;\Tr(X^{i_1}\cdots X^{i_L})\Tr(Y^{j_1}\cdots Y^{j_L})
e^{-S_{\rm 2pt}},\cr
&S_{{\rm 2pt}}=\Tr(X^mY^m)+{g^2\o16\pi^2}\Tr[X^m,X^n][Y^m,Y^n]
+{g^2\o32\pi^2}\Tr[X^m,Y^n][X^m,Y^n]~.
}}
Similarly, the three-point function coefficients are reproduced by
the  matrix model interaction
\eqn\Sthreept{\eqalign{
S_{{\rm 3pt}}^{(int)}=\,2\Tr[X^m,X^n][Y^m,Z^n]&+\Tr[X^m,Y^n][X^m,Z^n]\cr
+2\Tr[Y^m,Y^n][Z^m,X^n]&+\Tr[Y^m,Z^n][Y^m,X^n]\cr
+2\Tr[Z^m,Z^n][X^m,Y^n]&+\Tr[Z^m,X^n][Z^m,Y^n]~.
}}

\listrefs

\end